\documentclass[preprint]{emulateapj}
\usepackage[normalem]{ulem}
 
\slugcomment{Submitted to {\it{The Astrophysical Journal}}}
\shortauthors{Gonzalez et al.}
\shorttitle{Galaxy Cluster Baryon Fractions}

\mathchardef\myhyphen="2D
\newcommand{\msun}{M$_\odot$}
\newcommand{\lsun}{L$_\odot$}

\newcommand{\kms}{~km~s$^{-1}$}

\newcommand\rfive{$r_{500}$}
\newcommand\rtwo{$r_{200}$}
\newcommand\pone{GZZ05}
\newcommand\ptwo{GZZ07}
\newcommand\mfive{${M}_{500}$}
\newcommand\mtwo{M$_{200}$}

\newcommand{\hst}{\textit{HST}}
\newcommand{\xmm}{\textit{XMM}}
\newcommand{\chandra}{\textit{Chandra}}

\newcommand{\wfc}{WFC3}

\newcommand{\mlrat}{$\Upsilon_\star$}
\newcommand{\mlrati}{$\Upsilon_{\star,I}$}

\begin{document}
\title{Galaxy Cluster Baryon Fractions Revisited \footnotetext[1]{Based on observations obtained with {\it XMM-Newton}, an ESA science mission with instruments and contributions directly funded by ESA Member States and NASA.}
}
  
\author{Anthony H. Gonzalez,\altaffilmark{1} 
 Suresh Sivanandam,\altaffilmark{2,3} Ann I. Zabludoff,\altaffilmark{2} \&
Dennis Zaritsky\altaffilmark{2}}
\altaffiltext{1}{Department of Astronomy, University of Florida, Gainesville, FL 32611-2055}
\altaffiltext{2}{Steward Observatory, University of Arizona, 933 North Cherry Avenue, Tucson, AZ 85721}
\altaffiltext{3}{Dunlap Fellow, Dunlap Institute, University of Toronto, 50 St. George St, Toronto, ON, Canada}
    
\begin{abstract}

  We measure the baryons contained in both the stellar and hot gas
  components for twelve galaxy clusters and groups at $z\sim0.1$ with
  $M=1-5\times10^{14}$ \msun.  This paper improves upon our previous
  work through the addition of {\it XMM-Newton} X-ray data, enabling
  measurements of the total mass and masses of each major baryonic
  component --- intracluster medium, intracluster stars, and stars in
  galaxies --- for each system.  We recover a mean relation for the
  stellar mass versus halo mass, $M_{\star}\propto
  M_{500}^{-0.52\pm0.04}$, that is 1$\sigma$ shallower than our
  previous result.  We confirm that the partitioning of baryons
  between the stellar and hot gas components is a strong function of
  \mfive; the fractions of total mass in stars and X-ray gas withing a
  sphere of radius \rfive\ scale as $f_{\star}\propto
  M_{500}^{-0.45\pm0.04}$ and $f_{gas}\propto M_{500}^{0.26\pm0.03}$,
  respectively.  We also confirm that the combination of the brightest
  cluster galaxy and intracluster stars is an increasingly important
  contributor to the stellar baryon budget in lower halo masses.
  Studies that fail to fully account for intracluster stars typically
  underestimate the normalization of the stellar baryon fraction
  versus \mfive\ relation by $\sim$25\%.  Our derived stellar baryon
  fractions are also higher, and the trend with halo mass weaker, than
  those derived from recent halo occupation distribution and abundance
  matching analyses.  One difference from our previous work is the
  weak, but statistically significant, dependence here of the total
  baryon fraction upon halo mass: $f_{bary}\propto
  M_{500}^{0.16\pm0.04}$.  For $M_{500}\ga2\times10^{14}$, the total
  baryon fractions within \rfive\ are on average 18\% below the
  Universal value from the seven year WMAP analysis, or 7\% below for
  the cosmological parameters from the Planck analysis. In the latter
  case the difference between the Universal value and cluster baryon
  fractions is less than the systematic uncertainties associated with
  the \mfive\ determinations.  The total baryon fractions exhibit
  significant scatter, particularly at $M_{500}<2\times10^{14}$
  \msun\ where they range from 60-90\%, or 65-100\%, of the Universal
  value for WMAP7 and Planck, respectively.  The ratio of the
  stellar-to-gas mass within \rfive\ ($M_{\star}/M_{gas}$), a measure
  of integrated star formation efficiency, strongly decreases with
  increasing \mfive.  This relation is tight, with an implied
  intrinsic scatter of 12\%.  The fact that this relation remains
  tight at low mass implies that the larger scatter in the total
  baryon fractions at these masses arises from either true scatter in
  the total baryon content or observational scatter in \mfive\, rather
  than late-time physical processes such as redistribution of gas to
  beyond \rfive.  If the scatter in the baryon content at low mass is
  physical, then our results imply that in this mass range it is the
  integrated star formation efficiency and not the baryon fraction
  that is constant at fixed halo mass.

\end{abstract}


\section{Introduction}
\label{sec:intro}

Precision measurements of the total baryon content test the degree to
which the expectation that clusters are nearly fair samples of the
matter content of the Universe is valid \citep{white1991}, and provide
insight on the key physical processes that may act to drive observed
total baryon fractions away from the Universal value \citep[e.g.,][and
references
therein]{nagai2007,mcnamara2007,mcnamara2012,mccarthy2008,mccarthy2011,simionescu2011}.
How the baryons are distributed amongst the various components in a
cluster also informs models for cluster assembly and cluster galaxy
evolution.  The partitioning of the baryons between the hot
intracluster medium and stars constrains the net, integrated efficiency with
which baryons are converted to stars
in massive halos (hereafter star formation efficiency) 
and clarifies the importance of halo mass
as a fundamental parameter tracing  star formation efficiency
\citep{bryan2000,lin2003}. This partitioning is also an important
consideration for cosmological tests that aim to use the gas fraction
to constrain $\Omega_M$ and $w$ \citep[e.g.,][]{allen2008}, because
variation will contribute to the observational scatter even for
massive clusters.  Finally, the partitioning of stellar baryons
between satellite galaxies, the brightest cluster galaxy (BCG), and
intracluster light provides a stringent test for models of cluster
galaxy evolution and cluster luminosity function evolution, as the
latter serves as a final reservoir for all stars liberated from other
cluster galaxies
\citep{conroy2007,purcell2007,behroozi2012,watson2012}.

In a pair of previous papers, we conducted an initial census of the
stellar baryon content in a sample of 24 nearby systems with prominent
BCGs \citep[][hereafter \pone\ and
\ptwo]{gonzalez2005,gonzalez2007}. In \ptwo\ we used this sample in
conjunction with published relations between \mfive\ and gas mass to
quantify the total baryon fractions on group to cluster scales.  Two
central results were that the total baryon fraction within \rfive\ is
roughly constant and $\sim20$\% below the Universal value from WMAP
\citep{komatsu2011}, and that the gas and stellar baryon fractions
have strong, opposite trends with system mass. The matching of the
falling stellar and rising gas mass fractions with \mfive\ supports
the suggestion made previously that the star formation efficiency
decreases with increasing system mass \citep{bryan2000,lin2003}. In
\ptwo\ we also emphasized the importance of including the intracluster
light (ICL).  Our observations demonstrated that the combination of
the brightest cluster galaxy and intracluster light (hereafter
BCG+ICL) dominate the stellar content within \rfive\ at group scales,
but contain a decreasing percentage of the total stellar mass with
increasing cluster mass. Omission of the ICL can thus bias both the
slope and normalization of the derived relation between stellar and
total mass.

\begin{deluxetable*}{llllllll}
\tablewidth{0pt}
\tablecolumns{8}
\tablecaption{X-ray Data Properties \label{dataproperty}}
\tablehead{
\colhead{Name} & 
\colhead{Obs ID\tablenotemark{1}} & 
\colhead{Date} &
\colhead{Type\tablenotemark{2}} & 
\colhead{Filter}&
\colhead{t$_{\mathrm{MOS}}$ (ks)} &
\colhead{t$_{\mathrm{MOS,ff}}$ (ks)}  &
\colhead{Aperture (\arcmin)} }
\startdata
Abell 0122  & 0504160101 & 2007 Dec 03 & GO & Medium & 55.9/56.0 & 48.5/49.4 & $1.1-3.6$\\
Abell 1651 & 0203020101 & 2004 Jul 01 & GO & Thin & 14.9/14.9 & 8.2/8.3 & $1.9-6.3$\\
Abell 2401 & 0555220101 & 2010 Oct 29 & S12 & Thin & 56.8/56.9 & 54.9/54.8 & $1.6-5.3$         \\
Abell 2721 & 0201903801 & 2005 May 13 &XSA & Thin & 17.5/17.6 & 9.5/9.9  & $1.3-4.3$\\
Abell 2811 & 0404520101 & 2006 Nov 28 &GO & Medium & 24.1/24.1 & 21.7/22.0  & $1.3-4.5$\\
Abell 2955 & 0555220201 & 2008 Aug 02 &S12  & Thin & 81.4/81.5 & 73.9/75.3  & $1.0-3.3$ \\
Abell 2984 & 0201900601 & 2004 Dec 27 & XSA & Thin & 28.8/28.8 & 27.9/27.5 & $0.9-3.0$\\
Abell 3112 & 0105660101 & 2000 Dec 24 & XSA & Medium & 23.1/23.1 & 22.4/22.5  & $1.7-5.8$\\
Abell 3693 & 0404520201 & 2006 Oct 14 & GO & Medium & 34.2/34.3 & 28.8/29.1 & $0.9-3.2$\\
Abell 4010 & 0404520501 & 2006 Nov 13 & GO & Thin & 19.4/19.4 & 17.7/18.3 & $1.3-4.3$\\
Abell S0084 & 0201900401& 2004 Dec 04 &XSA & Thin & 33.7/34.0 & 18.3/18.1 & $1.2-3.8$\\
Abell S0296 & 0555220301& 2008 Dec 26 &S12 & Thin & 68.8/68.9 & 51.2/53.6 & $1.3-4.5$\\
\hline
Abell 0478 & 0109880101 & 2002 Feb 15 &XSA & Thin & 69.7/125.0& 48.0/77.2 & $1.9-6.4$ \\
Abell 2029 & 0551780201 & 2008 Jul 17 &XSA & Thin & 46.0/46.0 & 32.7/33.4 & $2.4-8.1$ \\
Abell 2390 & 0111270101 & 2001 Jun 19 &XSA & Thin & 22.2/22.2 &  9.9/10.0 & $1.1-3.5$
\enddata
\tablenotetext{1}{XMM-Newton observation identification.}
\tablenotetext{2}{XSA denotes data from the XMM-Newton Science Archive; GO denotes data from our Guest Observer programs; S12 denotes systems from the Sanderson et al. (2012) program.}
\label{tab:basicinfo}
\end{deluxetable*}

\begin{figure*}
\epsscale{0.8}
\plotone{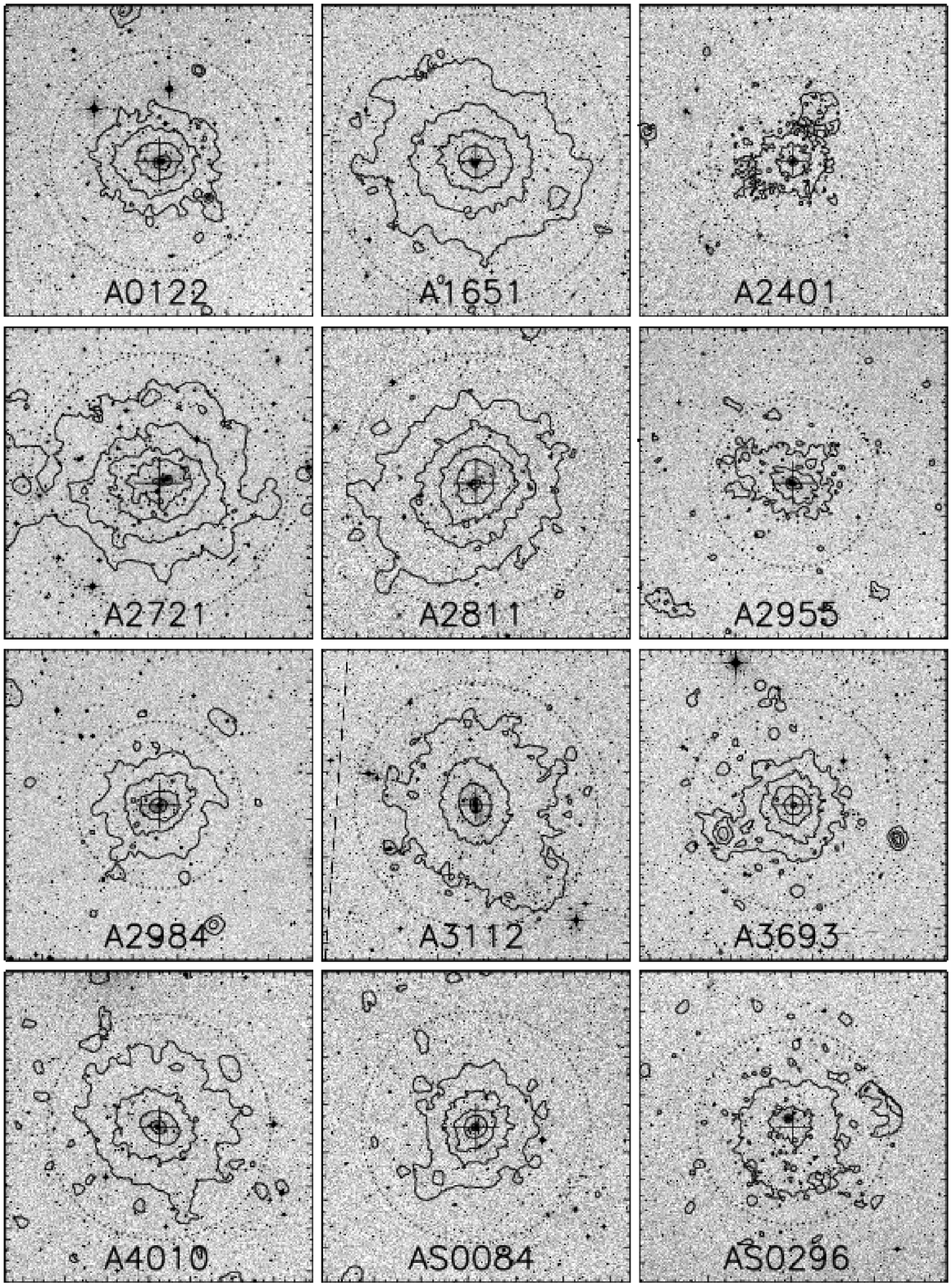}
\caption{$2.5\times2.5\:$Mpc thumbnails from the Digitized Sky Survey (DSS) for the primary sample of twelve clusters, with their X-ray surface brightness contours overlaid. The X-ray contours are generated from an adaptively smoothed surface brightness map. There are five contour levels which are logarithmically spaced at equal intervals. The lowest one corresponds to a value 3$\sigma$ above the brightness level measured at the outskirts of the {\it XMM} EPIC-MOS field-of-view, and the highest value corresponds to 0.1 dex below the brightest value. 
The crosshair indicates the fitted centre for the X-ray emission. The dotted circle is the measured $r_{500}$ value for each cluster.
 \label{fig:dssxrayplot}}
\end{figure*}

Similar results of declining stellar mass fraction with halo mass have
been obtained by a number of groups --- albeit in most cases excluding
the ICL.  Power law slopes for $f_*\propto M^{\alpha}$ in the
literature span the range $-0.65<\alpha<-0.25$ \citep[e.g.,
][]{lin2003,lagana2008,giodini2009,andreon2010,lagana2011,zhang2011,lin2012}. While
the case for a strong decrease of the stellar baryon fraction with
increasing halo mass is compelling, there exists disagreement among
these studies regarding the precise slopes of both the stellar and
total baryon fractions as a function of cluster mass.

A significant contributing factor to the present unsettled state of
the field is the quality of data available in past studies, including
our own. To conduct a comprehensive study of the baryon fraction
within a given radius (typically \rfive), the data should include deep
X-ray imaging for determination of the intracluster medium (ICM) gas
mass, redshifts to determine cluster membership far below $L^*$,
high-quality, multiband photometry suitable for both measurement of
the ICL contribution and stellar masses of cluster members, and a
reliable means of determining the cluster mass.  The only existing
study that satisfies all these criteria is \citet{sanderson2013},
which focuses upon a subset of the five lowest mass systems in the
current sample and thus lacks the dynamic range in mass to robustly
constrain the dependence of $f_\star$ upon \mfive.

Our previous study (\ptwo) had the advantage of directly measuring the
contribution of the ICL, a component that contains a significant
percentage of the stellar baryons and becomes increasingly important
with decreasing halo mass. However, a particularly critical limitation
of this work was the lack of X-ray data.  As a result, we constrained
the total baryon fraction only by combining the derived stellar baryon
fraction scaling relation with published X-ray data for a different
cluster sample. This limitation also led us in \ptwo\ to use cluster
velocity dispersion ($\sigma$) as a crude proxy for cluster mass.
Specifically, we used data from \citet{vikhlinin2006} to derive an
$M_X-\sigma$ relation to convert from dispersion to \mfive. The
uncertainty associated with this relation is relevant because, as
noted by \citet{balogh2008}, if our mass determinations are
systematically biased then the amplitude of the slope of the stellar
baryon fraction scaling relation will also be biased. This limitation
also precluded measuring scatter in the baryon fractions from cluster
to cluster variation.

The central aim of this paper is to improve upon \ptwo, directly
measuring the mass of each baryonic component plus the total cluster
mass for a subset of twelve galaxy clusters from \ptwo. As part of
this work, we also attempt to carefully assess the impact of any
systematic biases that may impact any of the observed measurements. In
section \ref{sec:sample} we describe the cluster sample and
data. Details of the total and gas mass measurements are provided in
section \ref{sec:totalandgasmass}, with a description of the stellar
luminosity and stellar mass measurements in section \ref{sec:stars}.
In section \ref{sec:results} we present our results for the
partitioning of stars between the BCG+ICL and the galaxies, the
partitioning of baryons between stars and gas, and the mass fractions
for each component. We also consider the impact of potential
systematics in this section and compare our results with previous
studies. Section \ref{sec:summary} presents the main conclusions from
this work.  Throughout the paper we adopt cosmological parameters
consistent with the \citet{komatsu2011} results from the seven-year
Wilkinson Microwave Anisotropy Probe data ($H_0=70.2$ \kms, $\Omega_M=0.275$,$\Omega_\Lambda=0.725$).
Changes between the seven- and nine-year WMAP cosmological parameters \citep{bennett2012}
have minimal effect on the results. We discuss the impact of the Planck cosmological parameters \citep{planckcosmology2013}
in section \ref{sec:planck}. We demonstrate that the main effect
of the change in cosmology is to reduce the offset between the cluster baryon fractions and the Universal baryon mass fraction.

\section{Cluster Sample}
\label{sec:sample}
The primary data set for this work consists of galaxy clusters drawn
from the \ptwo\ sample that have been observed by the {\it XMM-Newton}
X-ray telescope (hereafter \xmm). A total of ten clusters initially
satisfied this requirement. We exclude one of these clusters, Abell
3705, from our present analysis because the X-ray data indicate that
it is unrelaxed with multiple X-ray peaks \citep{sivanandam2009}. We
also include three additional systems from \ptwo\ that have recently
been targeted with \xmm\ by \citet{sanderson2013} to explore the
baryon fractions of low-mass systems, reanalyzing the data for
consistency with our other systems.

In addition to the twelve clusters in our primary sample, we also
analyze \xmm\ data for the most massive clusters from the
\citet{vikhlinin2006} sample to extend our mass baseline. While there
are many massive clusters with published results in the literature, we
opt here to only include ones from \citet{vikhlinin2006} that we have
reanalyzed to ensure that differences in analysis do not bias the
observed trends. Specifically, we reprocess \xmm\ data for the three
clusters from \citet{vikhlinin2006} with $T>7.5$ keV (Abell 478, Abell
2029, and Abell 2390).  While we lack the necessary data to constrain
the stellar masses, we use these clusters to place a lower limit on
the total baryon content at higher \mfive\ than are reached by our
primary sample.

\xmm\ observations for clusters presented in this work were either
taken through our guest observer (GO) programs, drawn from
\citet{sanderson2013}, or obtained from publicly available data in the
{\it XMM-Newton} Science Archive (XSA).  In Table \ref{dataproperty},
we present the details of the observations used in this work,
including the observation IDs, time of observation, data source,
filter choice, total integration time of the {\it XMM} EPIC-MOS camera
observations both before and after flare-filtering, and the aperture
used in the current analysis to derive cluster X-ray temperature.  We
present optical images of each cluster in our primary sample along
with their associated X-ray contours in Figure
\ref{fig:dssxrayplot}. The brightest cluster galaxy is nearly
co-centric with the peak of the X-ray emission, with offsets of less
than 75 kpc in all cases. A comparison of the outer (3$\sigma$)
contours to \rfive\ (dashed circles) illustrate that the data
typically reach beyond 0.7 \rfive\ before the cluster emission drops
below the background level.

\section{Total and Gas Mass from \xmm}
\label{sec:totalandgasmass}

We reduce the \xmm\ data using standard techniques for extended
sources. Our cluster observations generally fill a large portion of
the XMM field-of-view, requiring careful consideration of instrumental
and cosmic backgrounds.  For the processing we use \emph{XMM-Newton}
Science Analysis Software (SAS) version 12 with recent calibration
data from September 2012. We use the \emph{XMM-ESAS} software package
to carry out flare filtering and determine the quiescent particle
background \citep[see][for full details on \xmm\
backgrounds]{snowden2008}.  We restrict our attention to the MOS1 and
MOS2 cameras, as background estimation is more fully developed in
\emph{XMM-ESAS} for these than for the PN camera.

Our data analysis proceeds in two stages. The first stage consists of
spectral fitting to determine \rfive\ and \mfive. Given \rfive, we
then proceed to fit the surface brightness profiles for the \xmm\ data
to determine the gas mass profile and total gas mass within \rfive. We
provide a brief description of our full procedure below.

\subsection{Basic Processing}

The data are flare-filtered for soft proton flares using the
\emph{XMM-ESAS} flare filtering technique. The final integration times
after flare-filtering are listed in Table \ref{dataproperty}. In
general, greater than 50\% of the integration time of each cluster
observation is used. However, this culling may not remove low-level
soft proton flares, which need to be explicitly accounted for as part
of the analysis (see \S \ref{subsec:specfit}).  We next identify point
sources, which must be masked prior to spectroscopic or spatial
analysis of the extended cluster emission.  For source detection we
generate a combined $0.5-10$ keV joint MOS1 and MOS2 image and run the
wavelet detection algorithm \emph{ewavelet} with a 5$\sigma$ detection
threshold on scales of 4-8 pixels (10\arcsec$-$20\arcsec).  After
using \emph{emldetect} to determine the source extraction regions, we
visually inspect the source extraction regions to ensure that there
are no spurious source detections or sources associated with extended
cluster emission. In the few cases where the detection algorithm flags
cluster emission, we manually remove these detections from the point
source catalog.  This list is used to exclude regions in each
observation when extracting spectra or generating a surface brightness
map.  Finally, we note that we exclude data from individual chips in
instances where they exhibit anomalously elevated backgrounds in the
$0.5-10$ keV image relative to the other chips. The excluded chips
were typically CCD \#4 in MOS1 and CCD \#5 in MOS2.

\begin{figure}
\plotone{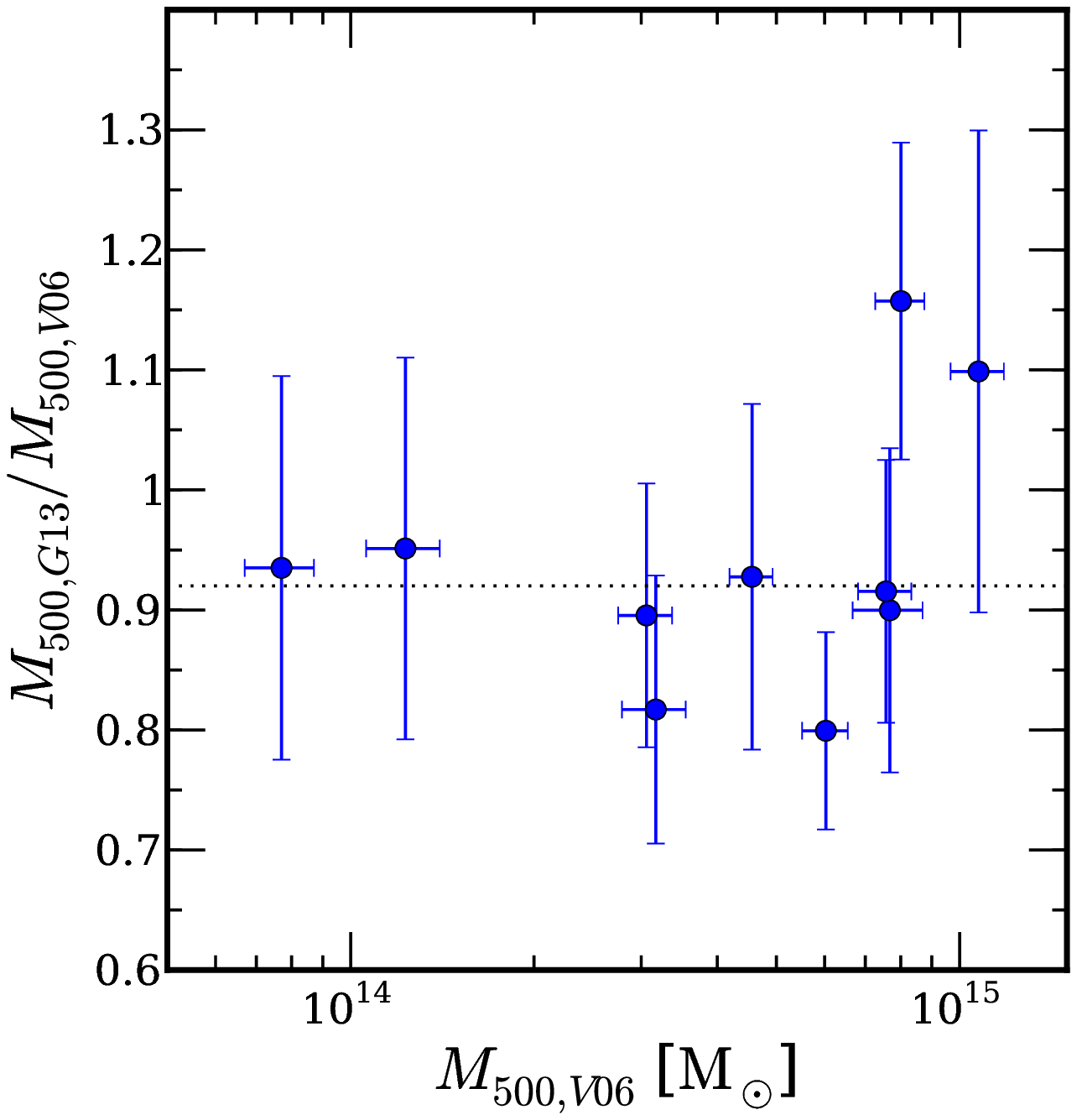}
\caption{
Comparison of \mfive\ measurements for a common set of clusters as measured by 
\citet[][V06]{vikhlinin2006} using \chandra\ data with our \xmm -derived values (G13).
Our \mfive\ values 
are on average 8\% below those of \citet[][dotted line]{vikhlinin2006},
with no observable dependence upon cluster mass.
 \label{fig:clustercomp}}
\end{figure}

\subsection{Spectral Fitting and $M_{500}$}
\label{subsec:specfit}

Because the sample is heterogenous in terms of depth, we choose a
straight-forward and consistent method for determining \mfive, which
simply requires that the signal-to-noise be sufficient to perform
spectral fitting out to at least 0.5\rfive.  We use the
\citet{vikhlinin2009} calibration to convert from X-ray temperature
measured within an aperture spanning 0.15 to 0.5 \rfive\ to the total
\mfive\ for each cluster.  The relevant equations from
\cite{vikhlinin2009} are
\begin{eqnarray}
T_X/T_{X,2} &=& 0.9075 + 0.00625 T_{X,2} \\
M_{500} &=& M_0(T_X/5\:\textrm{keV})^\alpha E(z)^{-1}
\end{eqnarray}
where $T_X$ and $T_{X,2}$ are the X-ray temperatures measured at
\rfive\ and 0.5\rfive, respectively, $M_0 =
(3.02\pm0.11)\times10^{14}h^{-1}M_\odot,$ $\alpha = 1.53\pm0.08,$ and
$E(z)$, which describes the evolution of the Hubble parameter, is
defined as in \citep[pp. 310-321]{peebles1993}.

Measuring $T_X$ is an iterative process where we first estimate
\rfive\ to compute the appropriate spectral extraction aperture, and
iterate until the computed value of \rfive\ is stable. The spectra are
extracted using the \emph{XMM-ESAS} software and fit using \emph{XSPEC
  12}.  \emph{XMM-ESAS} generates the appropriate quiescent particle
background spectrum, which is then subtracted from the spectra before
fitting. The fit model and method used for the spectral fitting is
outlined in \cite{snowden2008} and is identical to the one described
in our previous work \citep{sivanandam2009}. The fit function properly
accounts for instrument background such as residual soft proton flux
and fluorescent lines, the cosmic background, and the thermal emission
from the cluster ICM. Only the spectrum within the extracted aperture
is fit along with ROSAT All-Sky Survey X-ray background data to
constrain the cosmic background. The initial guess of \rfive\ is
computed using the temperature determined from an aperture with inner
and outer radii of $0\arcmin$ and $6\arcmin$, respectively.
The spectral extraction and fits are repeated until the inner and
outer extraction radii do not change by more than 10\arcsec. The
extraction aperture for each cluster are given in Table
\ref{tab:basicinfo} and the final derived $T_{X,2}$ are listed in
Table \ref{tab:data}.  The 1$\sigma$ uncertainties are determined by
carrying out 10,000 Monte Carlo simulations, which take into account
the uncertainties in $T_{X,2}$ and the \cite{vikhlinin2009} fit
functions. The derived \mfive\ values are also given in Table
\ref{tab:data}.

\subsection{Surface Brightness Profile Fitting and Gas Mass}                  

We compute the gas mass within \rfive\ using the gas density profile
determined from the imaging and spectral information. We use the same
methods described in \cite{sivanandam2009} with a few minor
refinements. The technique involves fitting a radially symmetric
projected $\beta$ profile \citep{cavaliere1978} to an X-ray image of
the cluster to determine the structural parameters of the gas density
profile, i.e., $r_0$ and $\beta.$ The central gas density is determined
using the cluster model's normalization (\texttt{norm}) parameter obtained from
the \emph{XSPEC} spectral fits within a given radial bin combined with
the structural parameters. One change from our previous analysis is
that we now use the $0.4-1.25$ keV X-ray images for the surface
brightness fits rather than a simultaneous fit to the $0.4-1.25$ keV
and $2-7.2$ keV images. As some of the clusters have cool-cores, this
approach minimizes the temperature dependence of the structural parameters. We
also model the instrumental background component slightly
differently. We include a flat instrumental background, which is set
to be the average quiescent particle background as determined by the
\emph{XMM-ESAS} software. The instrument background is held fixed
during the initial fit, which determines the projected $\beta$ model
parameters and the cosmic background contribution. After the initial
fit converges, a new fit is carried out where the value of the
instrument background is allowed to vary. This freedom is necessary to
account for residual soft proton flux not removed during flare filtering. With these final fit
parameters, the $\beta$ model is integrated out to \rfive\ to obtain
$M_{gas,500}$. Fit parameters and calculated properties are listed in
Table \ref{tab:data}.

\begin{figure}
\epsscale{1.0}
\plotone{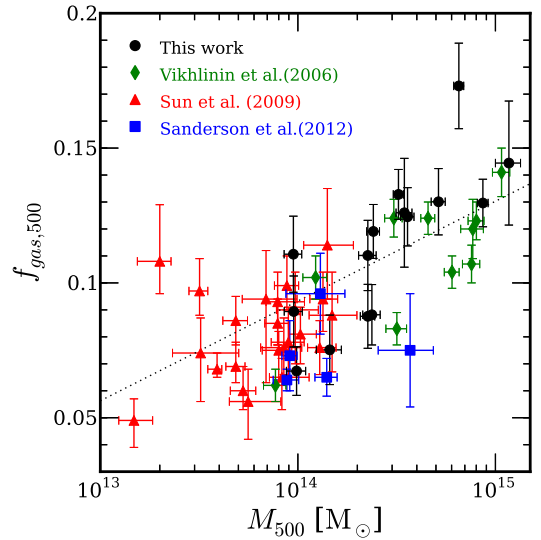}
\caption{X-ray gas fraction within $r_{500}$ as a function of $M_{500}.$ Different data sets are shown as a comparison with our work (black circles). The red triangles are from a {\it Chandra} study of groups \citep{sun2009}, while the green diamonds are {\it Chandra} measurements of clusters by \citet{vikhlinin2006}. We also include as boxes the measurements from \citet{sanderson2013} for five systems in our sample.
The dotted line is the fit to the gas fraction as determined by \cite{vikhlinin2009}.  The consistency of X-ray and optical centers, and regularity of the X-ray countours, argue that these systems are roughly relaxed, although small X-ray substructures are present in several.
At low mass our gas fractions span a similar range to those from \citet{sun2009}. At higher mass our measurements are consistent with the mean trend, but are $\sim10$\% higher than the \citet{vikhlinin2006} data. This offset is consistent with arising simply from the normalization difference for \mfive\ shown in Figure \ref{fig:clustercomp}.
\label{fig:fgascomp}}
\end{figure}

\subsection{Comparisons with other measurements}
\label{subsec:gascomp}
We next check the consistency of our \mfive\ and gas fraction
($f_{gas}$) measurements with other studies.  Because the
$M_{500}-T_X$ relation from \citet{vikhlinin2009} was calibrated using
\chandra\ data, and there may exist slight offsets between \xmm\ and
\chandra\ temperatures \citep{nevalainen2010}, we compare our \mfive\
with \mfive\ measured using \chandra\ data\citep{vikhlinin2006} for a
common set of clusters.{\footnote{ We note that while our \mfive\
    values are derived using an integrated temperature measurement,
    \citet{vikhlinin2006} use gas density and temperature profiles
    that extend out to \rfive\ to directly estimate hydrostatic masses
    for these systems within that radius.}}  The result of our
comparison is shown in Figure \ref{fig:clustercomp}.  While the
majority of clusters are consistent to within the statistical errors,
the \xmm\ masses are on average 8\% lower.  This offset, which is
likely the result of differences in the cross-calibration of \chandra\
and \xmm\ \citep{nevalainen2010}, provides an estimate of the
systematic uncertainty in \mfive\ from instrumental effects.
In section \ref{sec:gasbias} we also consider the potential systematic uncertainty
associated with the use of hydrostatic gas masses.

We also compare our $f_{gas}$ with other group and cluster samples
(Figure \ref{fig:fgascomp}). We compile these measurements from three
different samples, \citet{vikhlinin2006}, \citet{sun2009}, and
\citet{sanderson2013}. The two \chandra\ studies, \citep{sun2009} and
\citet{vikhlinin2006}, together cover the full mass range from groups
to clusters.  We find that our measurements are consistent with the
mean trend, but systematically offset to higher gas fractions by
$\sim$10\% relative to the \citet{vikhlinin2006} data (see
\ref{sec:gascomp}).  The observed offset is consistent with arising
simply from the normalization difference for \mfive\ shown in Figure
\ref{fig:clustercomp}. At low mass, our gas fractions span a similar
range to those from \citet{sun2009}.

\section{Luminosity and Stellar Mass Determinations}
\label{sec:stars}
\subsection{Luminosities}

For each component,
we compute the total luminosity within \rfive\, using the radius
derived from the \xmm\ data. 
To compute the luminosities of the BCG+ICL and cluster 
galaxy population, we use the same data and techniques as in \pone\ and \ptwo. As
described in \pone, all imaging was obtained in Gunn $i$, with photometry calibrated
to Cousins I using Landolt standards.
In converting from magnitudes to
luminosities, we take the absolute magnitude of the Sun to be
$M_I=4.09$ \citep[Vega;][]{mancone2012},{\footnote{Based upon the
    solar spectrum at
    http://www.stsci.edu/hst/observatory/cdbs/calspec.html}} which is
fainter than the normalization, $M_I=3.94$, used in \ptwo.  This
change in normalization results in all luminosities -- and hence
stellar masses -- increasing by 15\% relative to \ptwo, and 
corrects an inconsistency with \citet{cappellari2006} in our 
previous paper. 
   The
luminosities for the BCG+ICL component are computed using the same
models, 
and the resulting values are typically identical to those quoted in
Table 1 of that paper, modulo the change in $M_I$, because the
physical extent of this component is generally much smaller than
\rfive. Error bars are computed by propagating the uncertainties in
the model fit to the BCG+ICL component.

We compute the total luminosity of the galaxy population, excluding
the BCG+ICL, using statistical background subtraction. Specifically,
we compute the total flux from galaxies with $m_{BCG}<m_I<18$ within
\rfive, and use a $30^\prime - 60^\prime$ annular aperture encircling
the cluster to compute a statistical background correction for
subtraction. In \ptwo\ we assessed the potential impact of large-scale
structure on the derived background corrections via comparisons with
the SDSS luminosity function from \citet{blanton2003}, finding it to
be minimal.  We next convert from flux to absolute magnitude using the
cluster redshift, and apply a completeness correction to account for
galaxies with $m_I>18$. For this correction we adopt the cluster
luminosity function of \citet{christlein2003}, using $R-I=0.82$ to
convert between filters.  Additional details of this procedure are
discussed in \citet{gonzalez2007}.  The two dominant uncertainties in
the total galaxy luminosity are uncertainty associated with the
statistical background subtraction and systematic uncertainty
assocated with the completeness correction for the faint end of the
luminosity function. As noted in \ptwo, the derived luminosity is 12\%
lower if one assumes $\alpha=-1$ instead of $\alpha=-1.21$. No strong
dependence of the faint end slope upon mass is expected even at group
scales \citep[e.g.][]{zandivarez2011}, and so this assumption should
yield no scale-dependent bias. We therefore estimate a total
uncertainty of 0.15 mag for the integrated magnitude of the cluster
galaxy population.  The derived luminosities and associated
uncertainties for the BCG+ICL and the total luminosity of the system
are included in Table \ref{tab:data}.

\subsection{Stellar Mass}
\label{sec:stellarmass}

Conversion of observed luminosity to stellar mass requires choosing an
appropriate stellar mass-to-light ratio (\mlrat) for the cluster
galaxy population. While in principle straightforward, in practice
this is the least well constrained input when computing stellar baryon
fractions. There are several approaches that one can take for this
conversion, as we discuss in \S \ref{sec:stellarbias}.

We choose to base our estimate for \mlrat\ on the dynamical results
from \citet{cappellari2006}, as in \ptwo.  The strength of this
approach is that it avoids the use of stellar population models and
associated systematic uncertainties.  There are however several
limitations.  First, the population of galaxies with robust dynamical
measurements of central mass-to-light ratios is predominantly
comprised of quiescent galaxies, which have systematically higher
\mlrat\ values than the star-forming population. In \ptwo\ we ignored the
impact of the star-forming population in altering the global mass-to-light
ratio.  We retain in this paper the assumption of a purely passive
population, as in \ptwo, but assess the impact of this assumption in
\S \ref{sec:stellarbias}.  Second, in the \citet{cappellari2006} study
the derived dynamical masses within the effective radius were noted to
include $0-30$\% dark matter contributions, indicating that the
stellar \mlrat\ values are smaller by a corresponding amount.  While
the contribution of dark matter in \citet{cappellari2006} can be up to
30\% for individual systems, it is generally less for the slow
rotators that best represent the most massive galaxies, and contribute
substantially to the total baryon fraction.  In this paper we assume
an average contribution of 15\% dark matter, a change from \ptwo\ in
which we assumed a 0\% contribution. The total impact of these
limitations is expected to be minor, as we discuss further in \S
\ref{sec:stellarbias}.

\begin{deluxetable}{lccc}
\tabletypesize{\scriptsize}
\tablewidth{0pt}
\tablecaption{Derived Mass Fractions ($r<$\rfive)}
\tablehead{
\colhead{Cluster}  & \colhead{$f_{gas}$} & \colhead{$f_{stellar}$} &  \colhead{$f_{baryons}$}\\
\colhead{       }  & \colhead{         } & \colhead{             }  & \colhead{             } 
 }
\startdata
Abell 0122 & $0.088\pm.012$ & $0.024\pm.002$ & $0.112\pm.012$ \\
Abell 1651 & $0.130\pm.012$ & $0.013\pm.001$ & $0.143\pm.012$ \\
Abell 2401 & $0.089\pm.013$ & $0.028\pm.003$ & $0.118\pm.014$ \\
Abell 2721 & $0.126\pm.020$ & $0.017\pm.002$ & $0.143\pm.020$ \\
Abell 2811 & $0.125\pm.011$ & $0.013\pm.002$ & $0.138\pm.011$ \\
Abell 2955 & $0.067\pm.010$ & $0.030\pm.004$ & $0.097\pm.011$ \\
Abell 2984 & $0.111\pm.014$ & $0.041\pm.005$ & $0.152\pm.015$ \\
Abell 3112 & $0.133\pm.009$ & $0.022\pm.002$ & $0.155\pm.009$ \\
Abell 3693 & $0.110\pm.013$ & $0.023\pm.003$ & $0.133\pm.013$ \\
Abell 4010 & $0.119\pm.010$ & $0.023\pm.003$ & $0.143\pm.010$ \\
Abell S0084& $0.088\pm.011$ & $0.022\pm.003$ & $0.110\pm.011$ \\
Abell S0296& $0.075\pm.013$ & $0.020\pm.003$ & $0.095\pm.013$ \\
\hline
Abell 0478 & $0.173\pm.016$ &  ---            & ---             \\
Abell 2029 & $0.130\pm.009$ &  ---            & ---             \\
Abell 2390 & $0.144\pm.023$ &  ---            & ---             \\
\enddata
\tablecomments{The quoted stellar baryon fractions include a deprojection correction, as discussed in the text.}
\label{tab:baryonfracs}
\end{deluxetable}

\begin{deluxetable*}{lccccccccc}
\tabletypesize{\scriptsize}
\tablewidth{0pt}
\tablecaption{Observed Cluster Properties}
\tablehead{
\colhead{Cluster} &\colhead{$z$} &  $T_{X,2}$ & \colhead{$L_{BCG+ICL}$} & \colhead{$L_{Total}$}  &\colhead {\rfive}           & \colhead {\mfive} & \colhead {$M_{gas,500}$}       & \colhead {$M_{\star,2D,500}$} & \colhead{$M_{\star,3D,500}$} \\
\colhead{       } &\colhead{  }  &   \colhead{(keV)} & \colhead{($10^{12}$\lsun)}         & \colhead{($10^{12}$\lsun)}        &\colhead {(Mpc)} & \colhead {($10^{14}$ \msun)}   & \colhead {($10^{13}$ \msun)} & \colhead {($10^{13}$ \msun)} & \colhead{($10^{13}$ \msun)}
 }
\startdata
Abell 0122 & 0.1134 &  $3.65\pm0.15$     & $0.84\pm0.03$ & $ 2.57\pm0.16 $ & $0.89\pm .03$ & $2.26\pm.19$ &  $1.98\pm .21$ & $0.68\pm .04$ & $0.55\pm .03$ \\
Abell 1651 & 0.0845 &  $6.10\pm0.25$     & $0.87\pm0.09$ & $ 3.11\pm0.22 $ & $1.18\pm .03$ & $5.15\pm.42$ &  $6.70\pm .32$ & $0.82\pm .06$ & $0.65\pm .05$ \\
Abell 2401 & 0.0571 &  $2.06\pm0.07$     & $0.33\pm0.01$ & $ 1.30\pm0.09 $ & $0.68\pm .02$ & $0.95\pm.10$ &  $0.85\pm .09$ & $0.35\pm .03$ & $0.27\pm .02$ \\
Abell 2721 & 0.1144 &  $4.78\pm0.23$     & $0.57\pm0.01$ & $ 2.81\pm0.21 $ & $1.03\pm .03$ & $3.46\pm.32$ &  $4.36\pm .57$ & $0.74\pm .06$ & $0.57\pm .04$ \\
Abell 2811 & 0.1079 &  $4.89\pm0.20$     & $0.85\pm0.14$ & $ 2.13\pm0.18 $ & $1.04\pm .03$ & $3.59\pm.28$ &  $4.47\pm .17$ & $0.56\pm .05$ & $0.47\pm .04$ \\
Abell 2955 & 0.0943 &  $2.13\pm0.10$     & $0.60\pm0.03$ & $ 1.33\pm0.08 $ & $0.68\pm .04$ & $0.99\pm.11$ &  $0.66\pm .05$ & $0.35\pm .02$ & $0.30\pm .02$ \\
Abell 2984 & 0.1042 &  $2.08\pm0.07$     & $0.86\pm0.03$ & $ 1.72\pm0.08 $ & $0.67\pm .01$ & $0.95\pm.10$ &  $1.05\pm .08$ & $0.46\pm .02$ & $0.39\pm .02$ \\
Abell 3112 & 0.0750 &  $4.54\pm0.11$     & $0.93\pm0.05$ & $ 3.33\pm0.23 $ & $1.02\pm .02$ & $3.23\pm.19$ &  $4.29\pm .16$ & $0.88\pm .06$ & $0.70\pm .04$ \\
Abell 3693 & 0.1237 &  $3.63\pm0.20$     & $0.71\pm0.07$ & $ 2.42\pm0.18 $ & $0.90\pm .03$ & $2.26\pm.23$ &  $2.49\pm .15$ & $0.64\pm .05$ & $0.51\pm .04$ \\
Abell 4010 & 0.0963 &  $3.78\pm0.13$     & $0.81\pm0.12$ & $ 2.65\pm0.21 $ & $0.92\pm .02$ & $2.41\pm.18$ &  $2.87\pm .11$ & $0.70\pm .06$ & $0.56\pm .05$ \\
Abell S0084& 0.1100 &  $3.75\pm0.20$     & $0.70\pm0.02$ & $ 2.46\pm0.16 $ & $0.91\pm .03$ & $2.37\pm.24$ &  $2.09\pm .16$ & $0.65\pm .04$ & $0.52\pm .03$ \\
Abell S0296& 0.0696 &  $2.70\pm0.21$     & $0.57\pm0.01$ & $ 1.30\pm0.07 $ & $0.78\pm .04$ & $1.45\pm.21$ &  $1.09\pm .10$ & $0.35\pm .02$ & $0.29\pm .01$ \\
\hline
Abell 0478 & 0.0881 &  $7.09\pm0.12$     &  ---            &  ---             & $1.28\pm 0.03$ & $6.58\pm0.38$ &  $11.5\pm 0.8$ & ---  &  --- \\
Abell 2029 & 0.0773 &  $8.41\pm0.12$	    &  ---            &  ---             & $1.42\pm 0.03$ & $8.71\pm0.55$ &  $12.0\pm 0.4$ & --- &   --- \\
Abell 2390 & 0.2329 &  $10.6\pm0.8$	    &  ---            &  ---             & $1.50\pm 0.07$ & $11.8\pm1.8$ &  $17.2\pm 1.0$ & --- &   --- \\
\enddata
\enddata
\tablecomments{Abell 0478, Abell 2029, and Abell 2390 are not part of the main sample. These clusters were included only in the X-ray analysis to extend the baseline to higher mass, but have no photometry equivalent to the other systems with which to measure the stellar mass. At the high mass end; however, the stellar component contributes a relatively small fraction of the total baryons. The luminosities include appropriate e+k corrections for each galaxy from \ptwo. The stellar masses are quoted
as observed, with no deprojection correction applied. }
\label{tab:data}
\end{deluxetable*}

\citet{cappellari2006} provide an empirical determination of \mlrat\
using Schwarzschild dynamical modelling of two-dimensional kinematic
data from SAURON.  Equation (9) in \citet{cappellari2006} quantifies
the luminosity dependence of \mlrat\ in the $I-$band, which is
consistent with more recent observations from ATLAS3D (Cappellari,
priv. comm).  We use a similar method as in \ptwo, computing a
luminosity-weighted \mlrati\ for $L>0.25L_*$ (the range over which the
SAURON relation is established) and using the same
\citet{christlein2003} luminosity function.  We then include a
correction for the estimated 15\% dark matter contribution, deriving
\mlrati$=2.65$. This value lies between the median \mlrat\ derived for Chabrier and Salpeter IMFs
via stellar population modelling in \citet{leauthaud2012},
  and is 26\% lower than the value used in \ptwo\ due to the combination of the dark matter correction
and a correction to our previous \mlrat\ calculation.  We note that
the resultant stellar masses are however only $\sim13$\% lower than
those used in our previous work because the dark matter correction is
offset by the change in the absolute magnitude of the Sun between the
two papers.  We use same the $e+k$ corrections as in \ptwo\ to account
for passive evolution between the redshifts of our clusters and that
of the \citet{cappellari2006} sample, which lie at
$z<0.01$.

One final consideration in deriving stellar mass is line of sight
projection. While the derived gas and total masses are
three-dimensional quantities, we measure the projected stellar mass
within \rfive.  In most previous studies, including \ptwo, no attempt
was made to deproject the stellar mass and derive an estimate of the
total stellar mass enclosed within a three-dimensional sphere of
radius \rfive. {\footnote{An exception is \citet{giodini2009}, in which the
authors do apply a deprojection correction.}}   For the current paper we compute both the observed and
deprojected stellar masses, hereafter $M_{\star,2D}$ and
$M_{\star,3D}$, respectively. We calculate $M_{\star,3D}$ using a
similar approach to \citet{sanderson2013}.  Specifically, we model the
galaxy distribution using an NFW profile with $c=2.9$
\citep{lin2004a}. For this concentration, 71\% of the galaxy stellar
mass within a projected \rfive\ lies within a sphere or radius
\rfive. We apply no deprojection correction to the BCG+ICL, as the
observed physical extent of the ICL is significantly less than \rfive\
in all cases. This deprojection correction decreases the normalization
of the stellar baryon fraction relation by $\sim20$\%. In addition,
because the importance of the galaxy population relative to the
BCG+ICL increases with \mfive, this correction has the greatest impact
at high mass and also acts to steepen the relation.  We list both the
projected and deprojected stellar masses in Table \ref{tab:data}.

\section{Results and Discussion}
\label{sec:results}
\subsection{Stripping Efficiency: Partitioning of Stars between the BCG+ICL and Galaxies}

\begin{figure}
\plotone{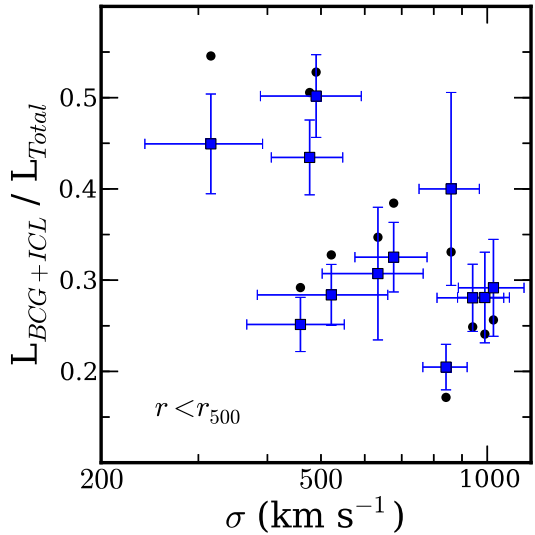}
\caption{Contribution of light within \rfive\ contained by the BCG+ICL as function of cluster velocity dispersion. The black circles denote the values from \ptwo; the blue squares correspond to the revised contribution based upon the new \rfive\ values from our current work, including the groups from Sanderson et al. (2012).}
\label{fig:bcgiclvd}
\end{figure}

Given the updated \mfive\ and \rfive\ determinations, we first
investigate what percentage of the cluster luminosity is contained
within the central BCG and ICL as a function of \mfive. In \ptwo\ we
found a strong trend of the BCG+ICL contributing a decreasing
percentage of the total cluster luminosity with increasing cluster
velocity dispersion. In Figure \ref{fig:bcgiclvd} we show an updated
version of this relation, comparing the original results with
luminosity fractions calculated within the \rfive\ radii derived from
the X-ray data.  The main change is a modest flattening of the trend,
as the percentages have decreased for the lower dispersion systems and
increased for the highest mass systems.  A fit to the data still
however yields $L_{BCG+ICL}/L_{Total}\propto \sigma^{-0.78\pm0.12}$
for $r<r_{500}$ within the range of dispersions covered by this
sample.

\begin{figure}
\plotone{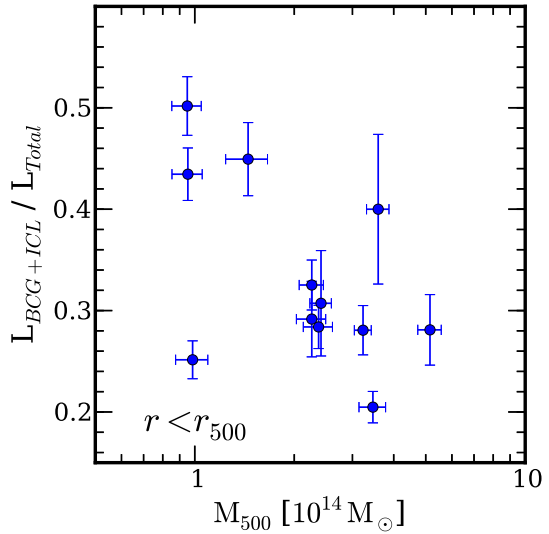}
\caption{Fraction of light within \rfive\ contained by the BCG+ICL as function of cluster mass. For the majority of low mass systems the BCG+ICL contribution to the total luminosity is nearly comparable to that of the galaxy population.  }
\label{fig:bcgiclfrac}
\end{figure}

With the inclusion of the X-ray data, we can now also move beyond
velocity dispersion, which is a high-scatter proxy for cluster mass,
and directly investigate the mass-dependence of the fractional BCG+ICL
contribution.  In Figure \ref{fig:bcgiclfrac} we plot the proportion
of the total luminosity within \rfive\ that is contained in the
BCG+ICL as a function of \mfive. 
The trend in fractional BCG+ICL contribution originally observed with
velocity dispersion in \ptwo\ is seen with \mfive\ at comparable
statistical significance, with $L_{BCG+ICL}/L_{Total}\propto
M_{500}^{-0.37\pm0.06}$.  This trend has also been confirmed
independently by searches for intracluster suernovae
\citep{mcgee2010,sand2011}.  Low mass systems in our sample exhibit a
substantially higher percentage of stars in the BCG+ICL than the most
massive systems, with $40-50$\% of the stellar luminosity contained in
this component for \mfive$\approx 1\times10^{14}$ \msun.  There is,
however, one outlier (A2955) at low mass and low BCG+ICL content from
the \citet{sanderson2013}
sample. While it is difficult to draw strong conclusions based upon a
single data point, this group may be indicative of significant scatter
in the partitioning of stars between BCG+ICL and galaxies at low mass.

\subsection{Star Formation Efficiency: Partitioning of the Stars and Gas}

\begin{figure*}
\plottwo{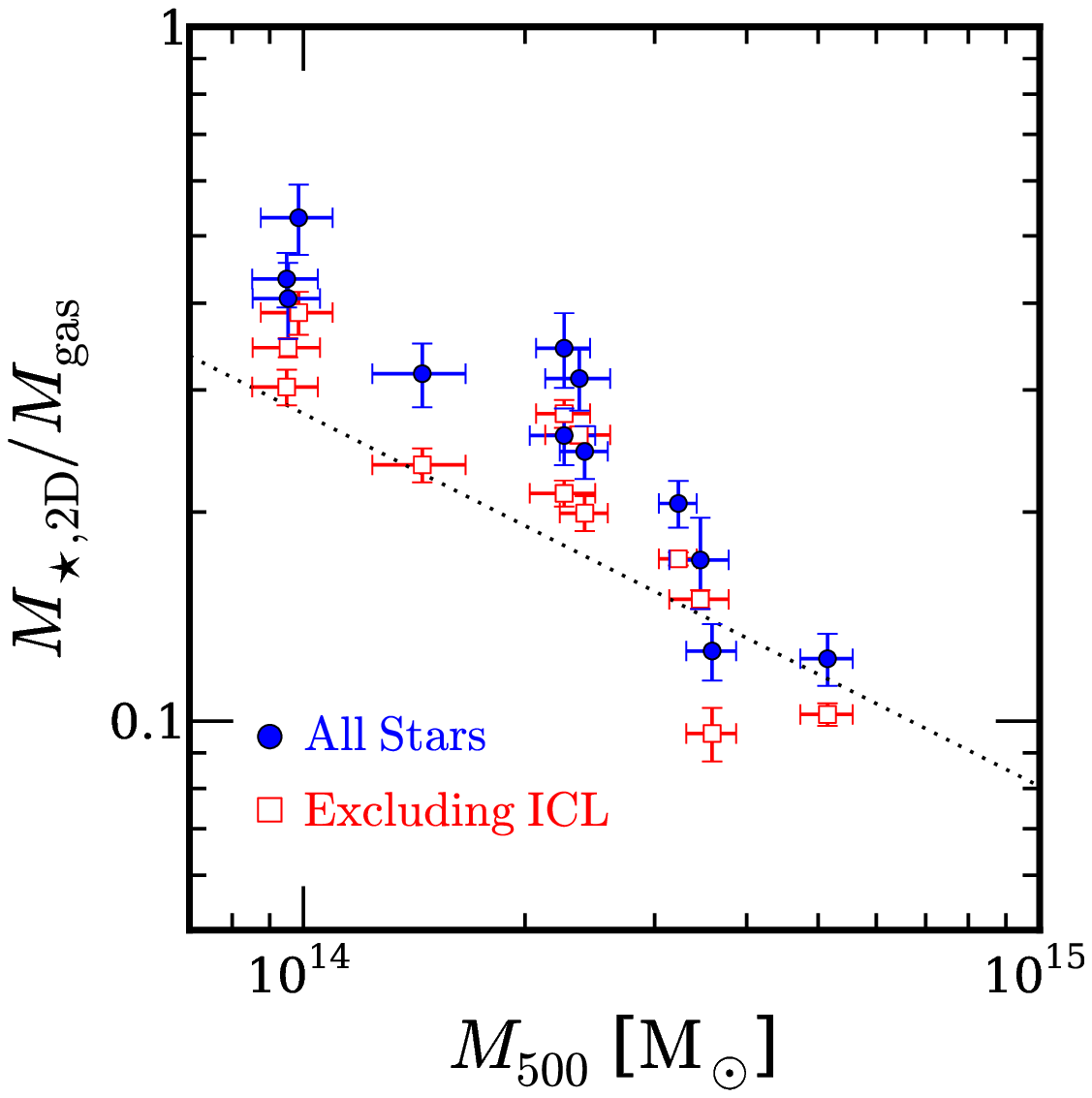}{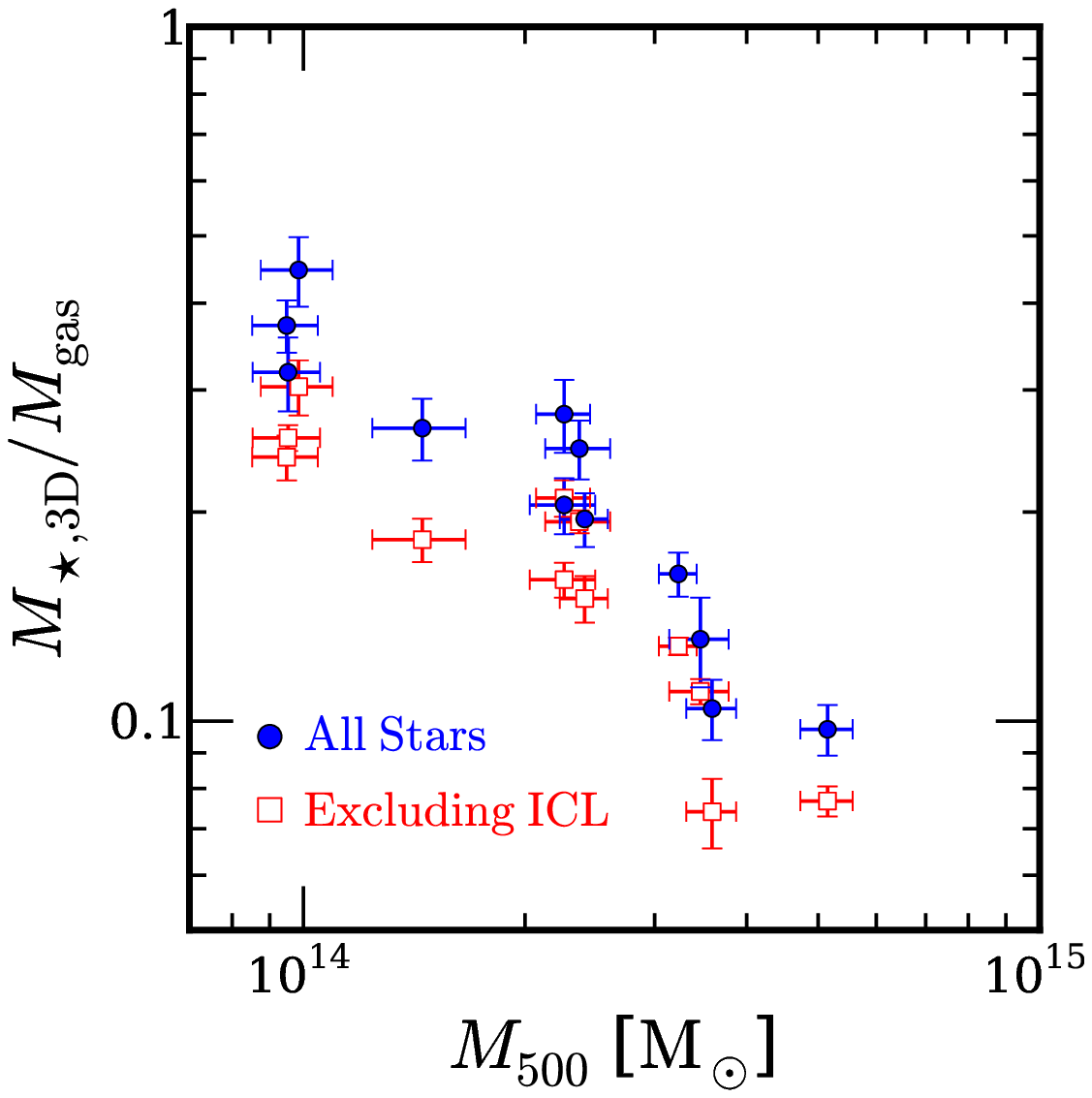}
\caption{Ratio of stellar to gas mass as a function of \mfive. The left panel shows the total stellar mass projected within \rfive, while the right panel includes a deprojection correction (see \S \ref{sec:sbf}). We plot this ratio for both the total stellar mass (BCG+ICL and cluster galaxies, blue filled circles) and the stellar mass contained only in the cluster galaxies and the BCG, excluding the ICL (red open squares). For comparison, we overplot as a dotted line the relation (with no deprojection correction) derived by \citet{zhang2011}, which did not include the ICL and assumed a Salpeter IMF in deriving stellar masses. \citet{zhang2011} also did not apply a deprojection correction, so the lefthand panel is most directly comparable. For the cluster galaxy population, our data yield a slope slightly steeper than of \citet{zhang2011}, while the contribution of the ICL steepens the slope such that at $M_{500}\approx10^{14}$ \msun\ the stars contribute $\sim40$\% as much mass as the gas within \rfive. 
  \label{fig:stargas}}
\end{figure*}

An important strength of the current data set is that it enables us to
compare the stellar and gaseous masses for individual systems, thereby
directly assessing how the efficiency with which baryons are converted
to stars depends upon cluster mass, as well as cluster-to-cluster
scatter about the mean trend. \citet{zhang2011} 
explore this question, albeit without inclusion of the ICL, using XMM
and ROSAT data in combination with SDSS photometry. The authors find a
relatively strong dependence, $M_{\star,2D}/M_{gas}\propto
M_{500}^{-0.537\pm0.101}$, within \rfive\ with an intrinsic (physical)
scatter of
$29\pm5$\%. In Figure \ref{fig:stargas} we present our measurement of
this ratio, along with a comparison to the best-fit relation from
\citet{zhang2011}, which assumed a Salpeter IMF and used separate
\mlrat\ for the star-forming and quiescent populations. For a program such
as \citet{zhang2011}, which uses SDSS photometry for the BCG, we
expect that $\sim60$\% of the total light in the BCG+ICL will be
missed.{\footnote{ \citet{gonzalez2005} find that a 50 kpc aperture
    contains $\sim40$\% of the total light in the BCG+ICL. Magnitudes
    measured within this aperture are close to the Kron total
    magnitudes derived for BCGs \citep[e.g.,][found consistency to
    within 0.05 mag]{stott2010}, so a 60\% fractional reduction in the
    total luminosity should be a reasonable estimate.}}  We therefore
also include points for which we have removed 60\% of the BCG+ICL
stellar mass in our systems to enable a more direct comparison of
slopes between the two data sets. We define this as the ``50 kpc''
case below.  Residual differences will be present in the
normalizations due to the different treatments of the IMF and star-forming
versus quiescent populations, as well as to known biases in SDSS
photometry for brightest cluster galaxies
\citep{lauer2007,bernardi2007,hyde2009}.{\footnote{It has been pointed
    out by multiple authors that the SDSS magnitudes, such as those
    used by \citet{zhang2011}, are underestimates due to
    oversubtraction of the sky in crowded regions and for extended
    sources \citep{lauer2007,bernardi2007,hyde2009}, with
    \citet{hyde2009} finding magnitude corrections of up to 30\%. We
    make no attempt here to exclude additional luminosity in our
    systems to match this effect in the SDSS photometry.}}

If we only include the stellar baryons in the cluster galaxy
population and the BCG+ICL contribution within the inner 50 kpc (red,
square data points), then we recover a slope, $-0.79\pm0.05$, which is
approximately 2.5$\sigma$ steeper than \citet{zhang2011}. Inclusion of
the ICL beyond 50 kpc slightly steepens this slope to $-0.78\pm0.04$
due to the larger fractional contribution of the ICL at lower mass.
The best fit relation is presented in Table \ref{tab:bestfit} along
with the best fit for $M_{\star,3D}/M_{gas}$, which has a comparable
slope.

The normalization of the relation confirms the claim in
\ptwo\ that the stellar component contributes nearly half of the
total baryonic mass by group scales.
The 17\% scatter about the best fit relation can be
explained by a combination of the known observational uncertainties and 12\%
intrinsic scatter. The physical implication of this low intrinsic scatter is that star
formation efficiency at a fixed cluster mass is quite uniform.

\subsection{Stellar, Hot Gas, and Total Baryon Mass Fractions}

The combination of X-ray and optical data described above enable us to
conduct the first analysis of the baryon mass fractions within \rfive\
for a statistical sample of clusters in which each system has direct
determinations of the stellar, intracluster, and dark matter
contributions within comparable radii.

\subsubsection{Gas Mass Fraction}
\label{sec:gascomp}
We fit the relation $M_{gas}=a M_{500}^b$, where $M_{gas}$ is the gas
mass within \rfive\, using orthogonal distance regression. We list the
best fit parameters in Table \ref{tab:bestfit}.  The gas fractions
derived in this paper yield a mass dependence slightly steeper than
the fiducial relation derived from \citet{vikhlinin2006} that was used
in \ptwo. The slope in \ptwo, which corresponds to $b-1$, was
$0.20\pm0.05$; here we report $0.26\pm0.03$.  The normalization of the
relation is also 8\% higher than the one derived from the data in
\citet{vikhlinin2006}, which is consistent with the offset in total
mass presented in Figure \ref{fig:clustercomp} and \S
\ref{sec:totalandgasmass}.  The uncertainty in the total 
masses due to both calibration and departures from hydrostatic equilibrium
is our largest source of systematic uncertainty in the gas fractions. As discussed in
section \ref{sec:gasbias}, the derived slope is robust to this uncertainty.

\subsubsection{Stellar Mass Fraction}
\label{sec:sbf}

While a steep decrease in stellar baryon fraction with increasing
cluster mass has been observed now by a number of groups \citep[e.g.,
GZZ07;][]{giodini2009,andreon2010,zhang2011,lin2012}, the most recent
generation of simulations have difficulty reproducing this trend
\citep[e.g.,][]{kravtsov2009,puchwein2010,young2011}. 
The much weaker mass dependence found in these simulations, when coupled with
their matching of the gas fraction versus halo mass relation, implies too weak a
dependence of star formation efficiency on \mfive.
Additionally, 
\cite{balogh2008} have argued
on analytic grounds that slopes steeper that $\alpha\sim-0.3$ are
difficult to reconcile with hierarchical structure formation in a
$\Lambda$CDM cosmology --- put simply, it is hard to form clusters
with low stellar baryon fractions through the assembly of lower mass
systems with higher stellar baryon fractions. In that study, the
authors postulated that the \ptwo\ results could potentially be
reconciled with theoretical expectations if the velocity dispersion
mass estimates were systematically biased for the lowest mass
systems. The improved, independent masses in our current work directly
address this concern.

\begin{figure}
\plotone{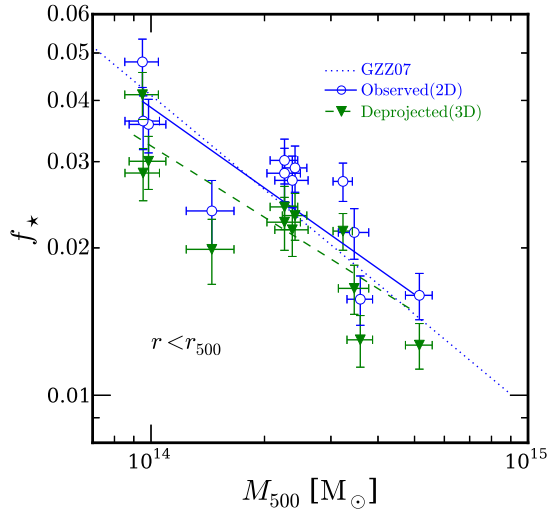}
\caption{ Stellar baryon fraction as a function of cluster mass for the systems in our sample. The
blue open circles denote the stellar mass determined within a projected radius of \rfive, and the solid line is
the best fit to the dependence of observed stellar mass upon \mfive\ over the range covered by the current
data. The dotted line is the best fit relation from \ptwo. Our current results confirm the trend in the
stellar baryon fraction from our previous paper. The green filled triangles correspond to the same data points after
applying a correction for projection effects to estimate the total stellar baryon content within
a sphere of radius \rfive\ centered on the brightest cluster galaxy. This correction lowers the inferred
stellar baryon fractions by $\sim20$ percent, with the best fit in this case denoted by the dashed line.
\label{fig:sbaryon}}
\end{figure}

We present in Fig. \ref{fig:sbaryon} updated data for stellar baryon
fractions as a function of \mfive.  We overplot the best fit relation
to the current data (solid line) compared with the best fit relation
from \ptwo\ (dotted).  In \ptwo\ we reported $\log f_{\star,2D}\propto
(-0.64\pm0.13)\log \mathcal{M}_{500}$. Here we find that a best
orthogonal distance regression fit to the full new set of data points
yields $\log f_{\star,2D}\propto (-0.45\pm0.04)\log
\mathcal{M}_{500}$, shallower than the previous value by 1.5 $\sigma$,
but still significantly steeper than $-0.3$.

One notable difference in the new stellar baryon fractions relative to
\ptwo\ is that the current data are less consistent with a simple
power-law fit with low scatter.  The few systems with
$M<2\times10^{14}$ have stellar baryon fractions that on average lie
below a simple extrapolation of the trend observed at higher mass. The
ratio of stellar to gas mass however shows little scatter or departure
from a pure power law over the full mass range.  Any departure from a
simple power-law must be driven by either lower true total baryon
fractions or by residual bias in the \mfive\ determinations. We return
to this point below in the context of total baryon fractions.

The overall normalization of the stellar baryon fraction relation
(prior to deprojection) is similar to that in our previous work.
While \mlrat\ is $\sim13$\% lower than in \ptwo\ (\S
\ref{sec:stellarmass}), the new \mfive\ values largely offset this
change.  The X-ray data used in this analysis enables us to avoid the
conversion from $\sigma$ to $T_x$ to \mfive\ via scaling relations as
in \ptwo, which is a clear improvement.

The application of a deprojection correction to estimate
$M_{\star,3D}$ is therefore the most significant change relative to
\ptwo. Because the BCG+ICL fraction decreases with \mfive,
deprojection slightly steepens the slope of the stellar baryon
fraction relation ($0.48\pm0.04$) in addition to decreasing the
overall normalization by $\sim20$\% (Figure \ref{fig:sbaryon}). We
list both the projected and deprojected stellar mass relations in
Table \ref{tab:bestfit}. We focus upon $M_{\star,3D}$ in the next
section when considering the total baryon content within \rfive.
The largest source of systematic
uncertainty in the stellar baryon fraction arises from uncertainty in 
the conversion from luminosity to stellar mass, which could change
the normalization of the derived relation by up to 15\% (see section \ref{sec:stellarbias}).
Uncertainty in the total mass is sub-dominant, having a minimal impact on the slope and
normalization (see section \ref{sec:gasbias}).

\begin{figure*}
\plotone{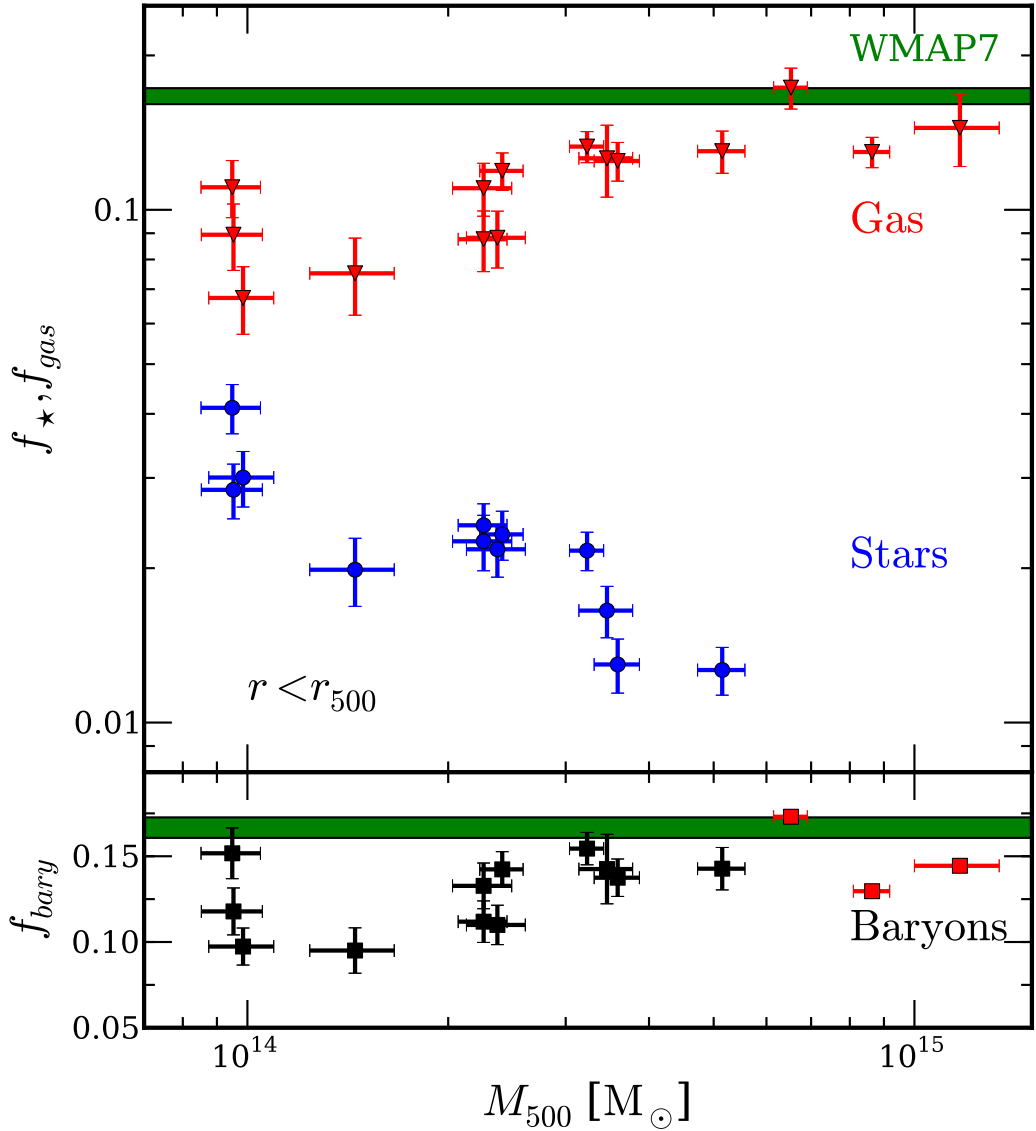}
\caption{Stellar, gas, and total baryon fraction as a function of cluster mass for the systems in our sample. The stellar
data (blue circles) include the deprojection correction, as discussed in the text.
Our best fit to this stellar data is $f_{\star,3D}\propto M_{500}^{-0.48\pm0.04}$, confirming evidence for a steep slope in our previous work. The gas fractions (red triangles) are plotted for all systems including the massive clusters from \citet{vikhlinin2006} for which we lack stellar fractions. The opposing trends for gas and stellar baryon fractions imply strongly decreasing star formation efficiency with increasing cluster mass. The total baryon fractions in the lower panel are plotted as black squares for all systems with stellar and gas data. We also include as red squares the gas fractions for the most massive clusters that lack stellar data; these are formally lower limits on the total baryon fractions, with stellar baryon fractions expected to be at the level of $f_{\star}\sim0.01-0.02$. The total baryon fractions are consistent with a weak, but statistically significant dependence of the total baryon fraction upon \mfive ($f_{bary}\propto M_{500}^{0.16\pm0.04}$) over the mass range where we have both stellar and gas measurements. This trend is driven primarily by the lower mass systems, for which we also see evidence of large scatter in the total baryon fractions.
The weighted mean baryon fraction for the current data at $M>2\times10^{14}$\msun, $f_{bary}=0.136\pm0.005$, is 18\% below the Universal value. } 
\label{fig:baryons}
\end{figure*}

\begin{figure*}
\plotone{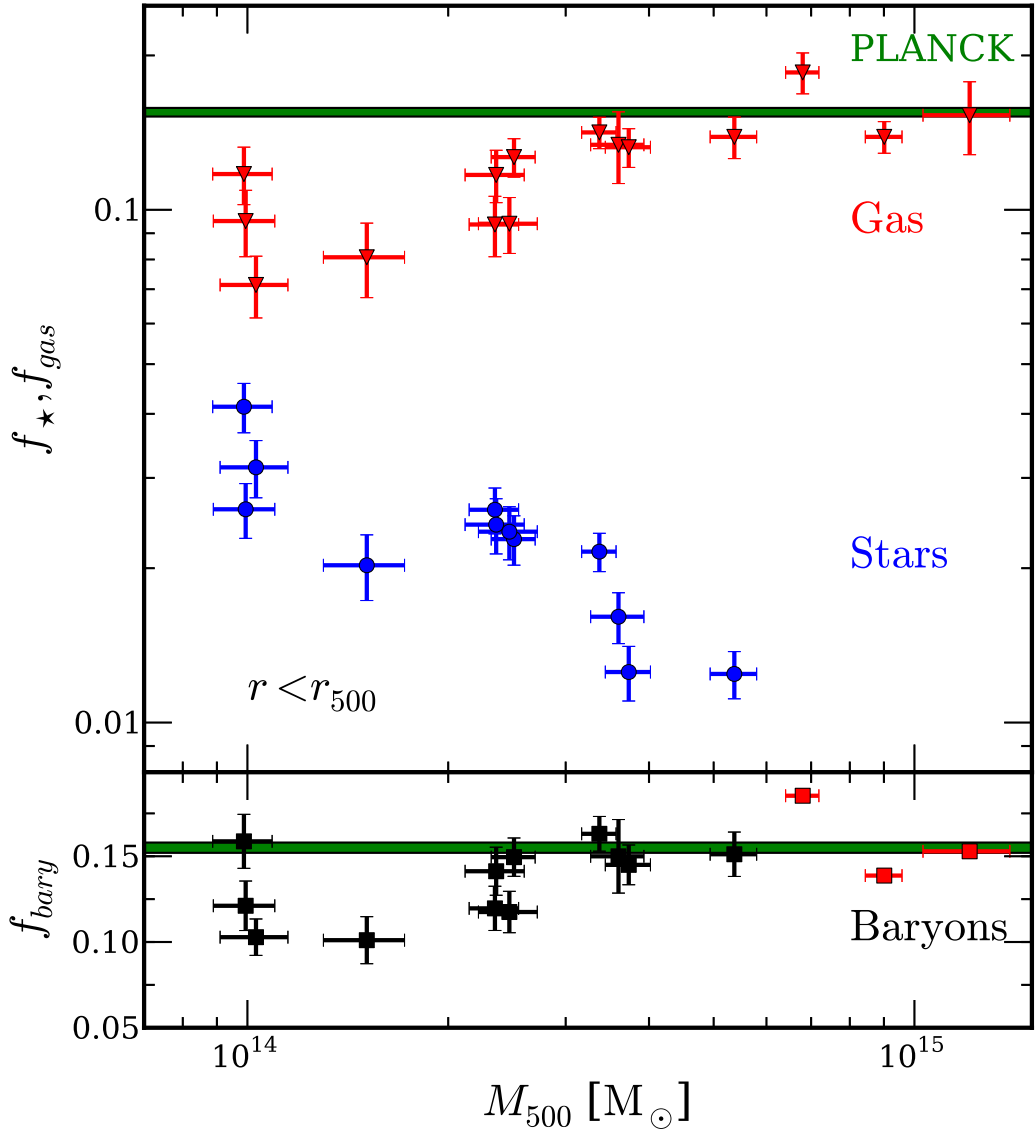}
\caption{Same as Fig. \ref{fig:baryons}, but using the Planck cosmological parameters from \citep{planckcosmology2013}. The best fit slopes to the
stellar and gas relations, which are provided in the Appendix, are minimally affected by the change in cosmology. The derived total baryon fractions however
 are closer to the Universal value when using the Planck cosmological parameters.
The weighted mean baryon fraction for the current data at $M>2\times10^{14}$\msun, $f_{bary}=0.144\pm 0.005$, is only 7\% below the Universal value. } 
\label{fig:planckbaryons}
\end{figure*}

\begin{figure}
\plotone{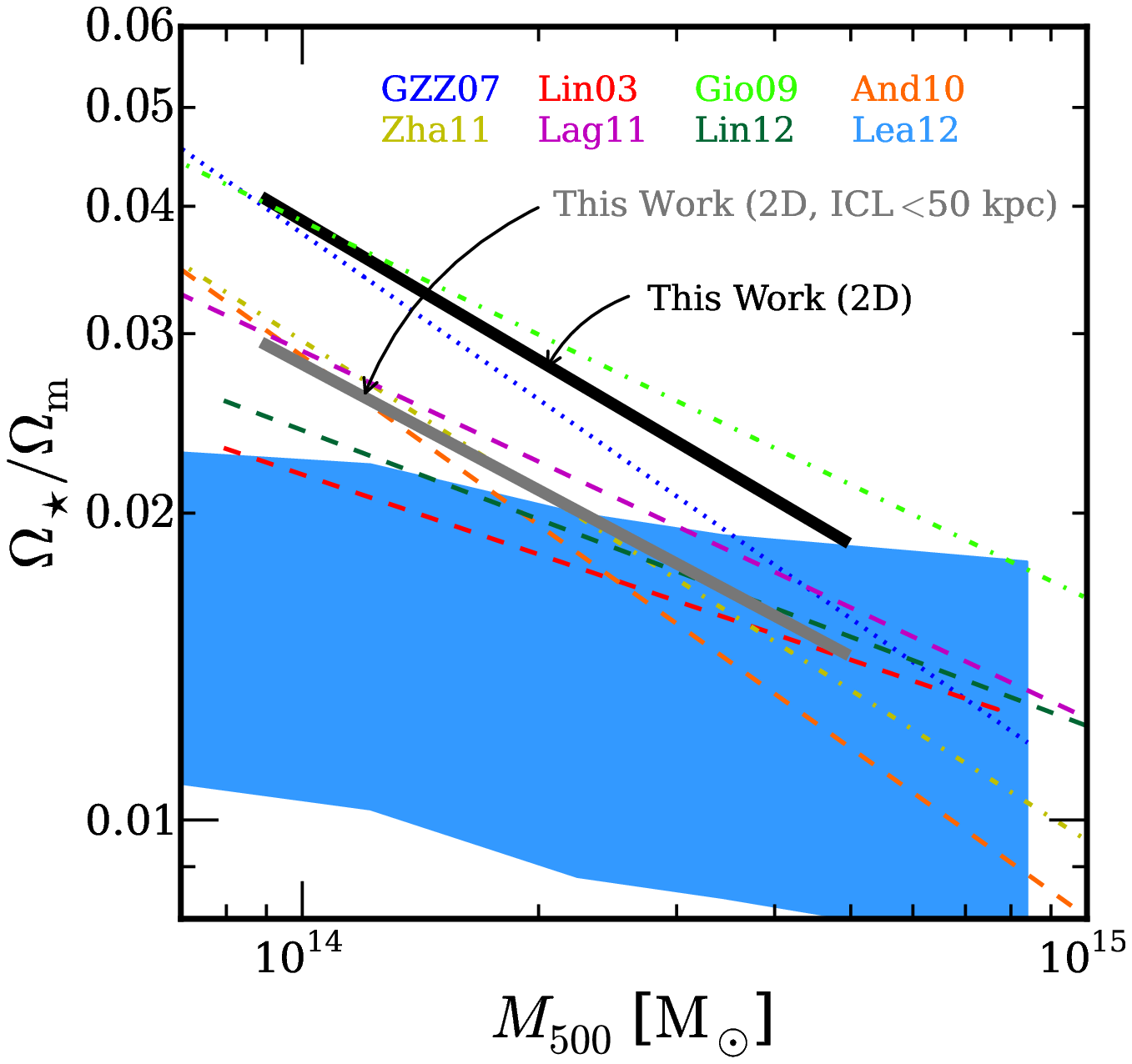}
\caption{
Comparison of the stellar baryon fraction relations in the literature with our observed relations, which illustrates
that our $\sim25$\% higher normalization is due to the significant contribution of ICL beyond 50 kpc, which is generally not included
in other studies. The increasing importance of ICL in low mass systems also acts to steepen the slope of the relation.
We  include published results from \citet[][Lin03]{lin2003}, \citet[][Gio09]{giodini2009}, \citet[][And10]{andreon2010}, \citet[][Zha11]{zhang2011}, \citet[][Lag11]{lagana2011}, \citet[][Lin12]{lin2012}, and \citet[][Lea12]{leauthaud2012}.  The \citet{andreon2010} relation is derived within \rtwo, while all others are within \rfive. The \citet{lin2003}, \ptwo, and \citet{andreon2010} mass-to-light ratios are based upon dynamical measurements, while the rest are shown for a Salpeter IMF. In the case of \citet{lin2012}, we have scaled the stellar mass up by a factor of 1.34 to convert from Kroupa to Salpeter IMF, while for \citet{leauthaud2012} we have used the shaded region corresponding to a Salpeter IMF from their Figure 5. For \ptwo\ we have changed the mass-to-light ratio to match that used in this paper.
The thick, solid black line is our observed relation within \rfive. If we approximately remove the contribution of the ICL beyond 50 kpc, we recover a relation similar in both slope and normalization to most other published studies within \rfive (``50 kpc'' case; thick, solid grey line).  If we apply a deprojection correction to our observed relation, the resulting normalization is similar to that of the relation for the ``50 kpc'' case
--- the contribution from the ICL at larger radii and the amplitude of the deprojection correction are roughly equal and opposite.  Both relations are inconsistent with the \citet{leauthaud2012} HOD results, particularly at low mass.
\label{fig:comp}}
\end{figure}

\subsubsection{Total Baryon Mass Fraction}

The total baryon mass fractions within \rfive\ for each individual
cluster are shown in the lower panel of Figure
\ref{fig:baryons} for a WMAP7 cosmology.
Above $2\times10^{14}$ M$_\sun$, the weighted mean baryon fraction for
the current data (including only systems with both stellar and gas
masses) is $f_{bary}=0.136\pm 0.005$, which is 18\% below the
Universal value. For comparison, in \ptwo\ we derived
$f_{bary}=0.133\pm 0.004$.  In \ptwo\ we found that the total baryon
fraction was consistent with being independent of cluster mass.  In
contrast, the current data are consistent with a weak dependence of
the baryon fraction upon cluster mass (Table \ref{tab:bestfit}).  The
derived power law slope is $0.16\pm0.04$.  The updated stellar masses
and deprojection correction have a minimal impact; the primary factor
driving this change is the addition of X-ray data for use in deriving
total and gas masses for individual systems.  In contrast, the overall
baryon fractions are higher, and the trend with mass weaker, than
found by \citet{leauthaud2012} using an extended halo occupation
distribution approach (HOD; see their Figure 11).  The similarity of
our derived slope with the result of
\citet[][$0.09\pm0.03$]{giodini2009} is likely coincidental. The
stellar mass fraction relation in that work, which does not include
the ICL, is high relative to most studies by an amount similar to our
estimated additional ICL contribution (Figure \ref{fig:comp}).

Systems with $M\la2\times10^{14}$ \msun\ show evidence for
larger scatter in $f_{bary}$ than do the higher mass systems, with
total baryon fractions ranging from $60-90$\% of the Universal value.
This large scatter, which is also discussed in \citet{sanderson2013},
is not however present in the ratio of stellar to gas
mass.\footnote{The baryon fractions presented in this paper are
  typically $\sim10$\% higher than those in \citet{sanderson2013} due
  to differences between the X-ray analyses, but the scatter is similar.}
Even if the total baryon fraction within a halo varies from the mean
expectation, the division of baryons between hot and cold components
is relatively unaffected.  Consequently, the observed scatter must be
due to either intrinsic variations in the total baryon content within
\rfive\ or remaining, unappreciated uncertainties in the derived
\mfive. While both the analyses in this work and in \citet{sanderson2013}
rely upon the same fundamental assumptions and include some of
the same systems, the fact that both observe large scatter for low
mass systems argues against a large bias in \mfive\ arising from details
of the analysis method. 
 
\begin{deluxetable}{lcr}
\tabletypesize{\scriptsize}
\tablewidth{0pt}
\tablecaption{Derived $M_j-$\mfive\ Relations}
\tablehead{
\colhead{Component}  & \colhead{$a$} & \colhead{$b$}
 }
\startdata
$M_{\star,2D}$& $3.9\pm0.2 \times10^{-2}$ &  $0.55\pm 0.04$\\  
$M_{\star,3D}$ & $3.2\pm0.1 \times10^{-2}$ &  $0.52\pm 0.04$\\ 
$M_{gas}$     & $ 8.8\pm0.3\times10^{-2}$ &  $1.26\pm0.03$\\   
$M_{bary}$   & $1.17\pm0.04\times10^{-1}$ &  $1.16\pm0.04$\\
\hline
$M_{\star,2D}/M_{gas}$  & $4.75\pm0.02\times10^{-1}$ &  $-0.82\pm0.05$\\
$M_{\star,3D}/M_{gas}$  & $3.90\pm0.02\times10^{-1}$ &  $-0.84\pm0.04$\\
\enddata
\tablecomments{ Best fit parameters for the relation $M_j=a (M_{500}/10^{14}M_\odot)^b$, where $M_j$ is the mass contained in each baryonic component. $M_{bary}$ is derived using the deprojected stellar mass. The slope for the baryon fraction relations is equivalent to $1-b$. We also include the best fit parameters for the stellar-to-gas mass ratios as a function of \mfive. The relation for the gas mass is derived including the clusters from \citet{vikhlinin2006}, while the other relations are derived using only clusters with both gas and stellar mass data.}
\label{tab:bestfit}
\end{deluxetable}

The existence of some systems at low mass with baryon fractions as
high as those of massive clusters raises the possibility that there
exists a flat upper envelope of baryon fractions across all masses. If
so, then the observed weak trend could simply be due to smaller
systems being increasingly susceptible to baryon loss and scattering
downward from their initial baryon fractions.  Conversely, it may be
that there is an intrinsic trend of $f_{bary}$ with mass, but with
increasing physical scatter with decreasing mass. We cannot currently
distinguish between these alternatives.

\subsubsection{Impact of Cosmological Parameters}
\label{sec:planck}
To assess the impact of the choice of cosmological model, we repeat our entire analysis
using the recently released cosmolgical parameters from Planck 
\citep{planckcosmology2013}
In Figure \ref{fig:planckbaryons} we show the gas, stellar, and total baryon mass fractions
derived using these parameters. The most striking difference relative to Fig. \ref{fig:baryons}
is the reduced offset between the cluster total baryon mass fractions and the Universal value.
Above $2\times10^{14}$ \msun, the weighted mean baryon fraction is $f_{bary}=0.144\pm0.005$, which
is only 7\% below the Universal value from Planck. Meanwhile, the change in parameters has minimal
impact on either the derived slopes (see Appendix), or on the scatter at lower mass, with total baryon fractions
ranging from 65-100\% of the Universal value from Planck.

\subsection{Potential Systematic Biases}
\label{sec:biases}

In the preceeding sections we showed that the BCG+ICL contribute
appreciably to the total stellar luminosity, especially for low mass
systems, confirmed that the stellar and gas fractions are strong,
inverse functions of \mfive, and demonstrated that the gas-to-stellar
baryon ratio is tightly correlated with \mfive. We also find that the
total baryon fraction is $\sim 17$\% (7\%) below the Universal values for
  systems with $M_{500}\ga 2\times10^{14}$\msun\ when using cosmological parameters
from the WMAP7 (Planck) analysis, with a weak but
statistically significant dependence upon \mfive. We now assess the
amplitudes of biases that may affect the stellar, gas, and total mass
determinations within \rfive, and consider the extent to which they
may alter our conclusions.

\subsubsection{Stellar Mass}
\label{sec:stellarbias}

For the stellar masses, the two stages in which biases might arise are
determination of the total luminosity and the conversion to stellar
mass.  As noted earlier, uncertainty in the faint end slope of the
galaxy luminosity function induces a 12\% uncertainty in the
integrated luminosity of the galaxy population, which corresponds to a
$5-8$\% uncertainty in the total luminosity including the BCG and
ICL. The luminosity of the intracluster light can also 
be underestimated if there is a more extended component at very low
surface brightness levels that is not co-centric with the BCG. We see
no evidence at higher surface brightness levels for such a component,
but include the possibility here for completeness.

A more immediate concern, as mentioned in \S \ref{sec:stellarmass}, is
the conversion of luminosity to stellar mass. There are several
approaches that one can take for this conversion. One approach is to
derive \mlrat\ based upon stellar population models using an assumed
spectral energy distribution.  This approach can either be done using
single-band photometry and a mean conversion for the ensemble galaxy
population \citep{lin2003,giodini2009,zhang2011,lagana2011}, or
preferrably using multiband photometry to fit for stellar masses for
each individual cluster galaxy \citep{leauthaud2012}. Deriving an
independent \mlrat\ for each galaxy has the advantage of making the
integrated results insensitive to variations in the passive galaxy
fraction from cluster to cluster.

All \mlrat\ estimates based upon stellar population models share a
common weakness, however, which is that they are only as accurate as
the underlying assumptions.  Specifically, the derived \mlrat\ is
strongly sensitive to assumptions about the shape of the initial mass
function and to systematic uncertainties in the stellar population
models themselves \citep[see e.g.,][for a detailed description of
these uncertainties]{conroy2009}, which can result in a significant
mis-estimation of the stellar baryon fraction. Moreover, recent
results from \citet{cappellari2012} argue that the initial mass
function is not universal, but rather is dependent upon galaxy mass
--- implying 
a different conversion than 
under the assumption of a universal IMF.

An alternative approach is to estimate 
\mlrat\ using dynamical measurements of the central mass-to-light
ratio for nearby galaxies. The dynamical approach, while less
sensitive to the IMF, is not entirely free of assumptions.  The
measured dynamical mass includes both stellar and dark matter
contributions, providing a direct observed upper limit on $M_\star$.
One must then estimate the fractional contribution of dark matter to
the dynamical mass in the central regions of galaxies in order to
estimate the stellar mass-to-light ratio. The standard technique for
estimating the dark matter fraction involves comparing the total
dynamical mass with the stellar mass one infers from stellar
population models. Using this technique, \citet{cappellari2006} found
that the dark matter contribution can be up to 30\% in individual
systems for a Kroupa IMF, but it is generally lower for slow rotators,
like the massive galaxies that contribute most to the total stellar
mass.

A number of recent studies find evidence for IMFs as steep as, or
steeper than, Salpeter in massive elliptical galaxies
\citep{conroy2012,vandokkum2012,cappellari2012}, and other stellar
systems \citep{zaritsky2012}. The average contribution of 15\% used in
this paper is chosen as the midpoint of values spanned by the range of
input assumptions. The associated systematic uncertainty is therefore
at most 15\% even considering the impact of the IMF. This remaining
uncertainty changes the slope of the relation between total baryon
mass and \mfive\ by $<1 \sigma$. Only if the IMF of cluster galaxies
were correlated with \mfive\ could one achieve a larger change in
slope.  Future results from the ATLAS$^\mathrm{3D}$ survey
\citep{cappellari2011} should further reduce this systematic
uncertainty.

The impact of the star-forming population upon the mean \mlrat, which we
have previously ignored, is a more subtle issue.  There are several
factors to consider. First, if the star-forming fraction depends upon
\mfive\ then this will induce a bias that scales with cluster mass, as
noted by \citet{leauthaud2012}. Recent studies yield somewhat
contradictory results for this dependence. Both \citet{finn2008} and
\cite{balogh2010} find that the fraction of star-forming galaxies is
nearly independent of mass from group to cluster scales, while
\citet{weinmann2006} and \citet{wetzel2012} conclude that the passive
fraction does increase with mass.  For the mass range under
consideration in this paper, $M>10^{14}$ \msun, the change is however
minimal.  For example, in \citet{wetzel2012} the passive fraction
increases by $\sim5$\% between $10^{14}$ \msun\ and $10^{15}$ \msun\
for massive cluster galaxies ($M_{stellar}\ga5\times10^{11}$ \msun),
with a $<10$\% increase even for the lowest mass clusters in that
study. If we assume the extreme, limiting case of \mlrat$=0$ for
star-forming galaxies, then a 10\% decrease in the passive fraction
changes the inferred \mlrat\ by only 4\% if the ICL contribution is
still assumed to be purely passive.

Second, the presence of star-forming galaxies in the cluster means
that we will systematically overestimate \mlrat\ because we are
assuming a purely quiescent population. To gauge the amplitude of
this effect for our data we use the formalism described in
\citet{lin2003}. As inputs to this calculation we take the true
passive fraction among cluster galaxies to be $\sim75$\%
\citep{balogh2010} and consider the BCG+ICL as an additional purely
passive stellar population, using the BCG+ICL fractions in
Fig. \ref{fig:bcgiclfrac}. We explore the impact of
three values of \mlrati\ for the star-forming population: 0.5, 1, and
2. If we weight by the relative contributions of the passive and
star-forming stellar populations then the composite \mlrati\ decreases
by 14\%, 12\%, and 7\%, respectively. Recently quenched
galaxies will also have lower \mlrat\ than galaxies whose star formation ended earlier.
The overall impact of this population is expected to be small given the rarity of
post-starburst galaxies in local clusters \citep[e. g.][]{couch2001,poggianti2004}.

Third, the \mlrat\ values in \citet{cappellari2006} are derived for a
sample of local galaxies. The galaxy clusters in our sample lie at
$z\simeq0.1$, and evolution since $z\sim0.1$ acts to increase \mlrat\
with time.  In our current analysis we make the approximation that
this evolution is passive, as in \ptwo. Any residual systematic
related to this evolutionary correction will be minor and subdominant
to other sources of systematic and statistical uncertainty.

Another concern is that in our analysis we have treated the BCG and
ICL as having the same \mlrat\ as the luminosity-weighted value for
the cluster population.  For the BCG, this approach likely
underestimates \mlrat. A BCG with a velocity dispersion of
$\sim300$\kms\ is expected to have \mlrati$\simeq4.8$ based upon
\citet{cappellari2006}.  In \pone\ we found that the BCG typically
contributes $\sim10-20$\% of the total light in the BCG+ICL, implying
a net \mlrat\ for the BCG+ICL $8-16$\% higher than the value assumed
here. Given that the fraction of light in the BCG+ICL is $<50$\%
(Figure \ref{fig:bcgiclfrac}), this corresponds to a potential bias of
$<10$\% for the total cluster stellar masses.

As for the ICL, which dominates the luminosity of the BCG+ICL, using
the same \mlrat\ as for the galaxy population is a reasonable
approximation if ICL predominantly originates from tidal stripping and
tidal disruption of cluster galaxies, consistent with theoretical
predictions
\citep{purcell2007,conroy2007,behroozi2012,watson2012,contini2013}.
Observational constraints on ICL color gradients support this picture:
the color of the ICL lies within the range found for the cluster
galaxy population \citep[][De Maio et al., in prep]{zibetti2005}.  The
\mlrat\ of the ICL would need to differ from the luminosity-weighted
\mlrat\ of the cluster galaxy population by more than 30\% for this to
be the dominant systematic uncertainty, an implausible level given
current observational constraints.

Finally, one can ask whether there is any systematic uncertainty
introduced by our methodology. As discussed in \S \ref{sec:stars}, our
approach has been to sum up the total luminosity, apply a background
correction, and then apply a single \mlrat\ weighted by a typical
cluster luminosity function.  An alternate approach is to compute the
stellar mass of each galaxy individually prior to the background
subtraction, assuming that it lies at the cluster redshift.

To assess the impact of the choice of methodology, we have repeated
our analysis with this alternate approach. For both the cluster and
background regions, we assign each galaxy an \mlrat\ according to its
magnitude using the \citet{cappellari2006} relation and assuming that
it lies at the cluster redshift, and then subtract off the background
contribution. As before, we apply a completeness correction to account
for the contribution of cluster galaxies fainter than the magnitude
limit of our data. For the BCG+ICL, we use \mlrat$=2.65$ as in or main
analysis. We again assume that all galaxies are quiescent, and refer
the reader to the discussion above on the impact of star-forming
galaxies upon the integrated stellar mass.  We find that this
alternate approach yields total stellar masses that are on average
15\% lower, corresponding to a systematic uncertainty comparable in
amplitude to the uncertainty associated with the dark matter
contribution in the \citet{cappellari2006} relation. The slope of the
stellar mass fraction relation meanwhile steepens by 1.5$\sigma$. For
the deprojected case it changes from $f_{\star,3D}\propto
M_{500}^{-0.48}$ to $f_{\star,3D}\propto M_{500}^{-0.54}$.

Considering all of the above factors, the largest
systematic uncertainties in our conversion to stellar mass are clearly
defined and impact our results at a level of $15$\%.  We therefore
consider this approach a robust alternative to the use of stellar population
models, which have much larger systematic uncertainty arising from the
IMF. For the total baryon fraction this corresponds to a maximum systematic uncertainty of $<5$\%, given that $M_\star/M_{gas}<0.5$ for all systems in our study.

A final consideration is potential bias arising from observing the
projected stellar mass within \rfive\ and applying a deprojection
correction to estimate the true stellar mass enclosed within a sphere
of radius \rfive.  For this correction we assume an NFW profile with a
concentration $c=2.9$ for the galaxy population \citep{lin2004a}, and
apply no correction to the more compact BCG+ICL component. As
discussed in \S \ref{sec:sbf}, this concentration implies that 71\% of
the stellar mass associated with the galaxy population within a
projected radius of \rfive\ lies within a sphere of radius
\rfive. There are several other estimates in the literature where the
concentration is measured in a similar fashion to
\citet{lin2004a}. \citet{carlberg1997} found $c\simeq3.7$ for CNOC
clusters, while \citet{budzynski2012} find $c\simeq2.6$ using a sample
of over 50,000 clusters and groups from the Sloan Digital Sky Survey.
Notably, \citet{budzynski2012} find that this concentration is
approximately independent of cluster mass over the mass range covered
by our study, indicating that any systematic uncertainty associated
with the mass dependence of the concentration is subdominant to the
uncertainty in the overall normalization.  The \citet{carlberg1997}
and \citet{budzynski2012} concentration values would give fractions of
73\% and 69\% of the projected stellar mass lying within a sphere of
radius \rfive, respectively. Thus, in the mean, the choice of
concentration value can induce a few percent bias in the stellar mass
associated with the galaxy population.  The impact on the total
stellar mass should thus be $\sim1-2$\%, depending on the contribution
of the BCG+ICL for a given system.

\subsubsection{Gas Mass and \mfive}
\label{sec:gasbias}

As noted in \S \ref{sec:gascomp}, our gas fractions are on average 8\%
higher than those derived by \citet{vikhlinin2006} for the same
systems. The gas fractions in general are offset by a similar amount
relative to the relation from \citet{vikhlinin2009} at high mass
($M_{500}\ga3\times10^{14}$ \msun), though they span a similar range
at lower mass as both the \citet{vikhlinin2009} and \citet{sun2009}
data.  The primary origin of this offset is expected be the 8\%
smaller \mfive\ values we derive from the \xmm\ data.  This remaining
systematic uncertainty in \mfive\ is one of the two dominant
systematic uncertainties in the current analysis (the other being
uncertainty in \mlrat).  As discussed in \S \ref{subsec:gascomp}, any
residual systematic error in \mfive\ is to first order expected to be
a constant fractional error independent of mass. It therefore should
only impact the overall normalization of the gas fractions.

A related concern is whether there exists any systematic bias endemic
in gas-based \mfive\ measurements due to departures from hydrostatic
equilibrium. The amplitude and direction of this systematic remains a
topic of significant ongoing research activity. While simulations by
\citet{lau2013} indicated that the bias should be at the few percent
level, recent studies comparing weak lensing and X-ray mass
determinations within \rfive\ do not yet unambiguously resolve this
issue. \citet{zhang2010} find minimal offset for their full cluster
ensemble, but a dichotomy between disturbed and relaxed clusters, with
the X-ray masses exceeding the lensing masses by 6\% for disturbed
systems and being 9\% low for relaxed systems. \citep{mahdavi2013}
meanwhile find that for their full sample the X-ray masses are low by
12\% relative to lensing, but with the cool-core systems having larger
relative X-ray masses.

We consider this potential bias associated with the hydrostatic
equilibrium approximation to be an outstanding issue that may impact
our \mfive\ determinations by up to $\sim15$\%. If, for example, our
derived \mfive\ (and hence \rfive) are too low then $M_{gas}$ and
$M_\star$ will also be underestimated due to their dependence on
\rfive.  For a 15\% increase in \mfive, $f_{gas}$ and $f_{\star}$
should decrease by $\sim8$\% and $\sim3$\%, respectively
\citep[c.f.][]{sanderson2013}.  The slopes of the gas and stellar
fraction relations are negligibly affected by such bias, regardless of
whether the masses are biased high or low.

One final potential concern is that our stellar mass measurements are
derived in an aperture centered on the BCG, while the gas and total
masses are derived within an equivalent aperture about the centroid of
the X-ray emission.  In practice, this difference in centering has a
minimal impact on our results. As can be seen in Figure
\ref{fig:dssxrayplot}, the offsets between the BCG and X-ray emission
are generally not large. For all but three clusters this offset is
less than 20 kpc, while the maximum offset is 75 kpc for Abell
S0296. As a consistency check, for the systems with the largest
offsets we recalculate the total baryon fractions with common centers,
finding that they change by $<1$\% in all cases.

\subsection{Comparison with Other Studies}
\label{sec:lit}

It is useful at this point to compare the stellar baryon fractions
derived in this work with 
the literature to assess the level of consistency among different
studies.  In particular, we consider results from \citet{lin2003},
\citet{giodini2009}, \citet{andreon2010}, \citet{lagana2011},
\citet{zhang2011}, \citet{lin2012}, and \citet{leauthaud2012}.  These
studies use a range of techniques to quantify the stellar baryon
fraction, each with different associated systematics, and thus the
ensemble provides an indication of the overall uncertainty in cluster
stellar baryon fractions. To enable as fair a comparison as
possible,
it is worthwhile to consider the key ways in which the methods differ.
\begin{deluxetable}{lcc}
\tabletypesize{\scriptsize}
\tablewidth{0pt}
\tablecaption{Best Fit Parameters For Baryon Fraction Relations}
\tablehead{
\colhead{Source}  & \colhead{Slope ($b-1$)} & \colhead{comments} \\
 }
\startdata
\citet{lin2003} & $-0.26\pm0.09$ &\\
\citet{lin2012} & $-0.29\pm0.04$ &\\
\citet{lagana2011} & $-0.36\pm0.17$ & \\
\citet{giodini2009} & $-0.37\pm0.04$ &\\
This paper & $-0.38\pm 0.05$ & No ICL at $r>50$ kpc (2D) \\
This paper & $-0.45\pm 0.04$ & Including full ICL (2D) \\
This paper & $-0.48\pm 0.04$ & Including full ICL (3D) \\
\citet{zhang2011} & $-0.49\pm0.09$ & \\
\citet{andreon2010} & $-0.55\pm0.09$ & Within \rtwo\\
\ptwo\ & $-0.64\pm0.13$ & Including full ICL (2D)\\
\enddata
\tablecomments{Slopes are sorted in order of decreasing value. All values are computed for \rfive\ except for \citet{andreon2010}.}
\label{tab:slopes}
\end{deluxetable}

There are three key issues. 
First, this paper and \ptwo\ are the only studies that directly
consider the contribution of the ICL to the stellar baryon content.
The other studies do include the BCG, either directly
\citep{lin2003,giodini2009,andreon2010,lagana2011,zhang2011,lin2012},
or by virtue of an extended HOD formalism \citep{leauthaud2012}. If we
make the rough approximation that the BCG magnitudes in these papers
correspond to the total light enclosed within the central 50 kpc, then
the BCG luminosities will include $\sim40$\% of the combined BCG+ICL
luminosity \citep{gonzalez2005}. For direct comparison with previous
studies we calculate the inferred stellar baryon fraction that we
would obtain if we were to use a similar approach. Excluding 60\% of
the BCG+ICL luminosity in this fashion yields a shallower stellar
baryon relation, $f_{\star,2D}\propto
M_{500}^{-0.38\pm0.05}$. 
This slope lies near the median of values derived in recent literature
within \rfive\, which range from $-0.26$ to $-0.49$ (see Table
\ref{tab:slopes} and Figure \ref{fig:comp}).

Second, \citet{andreon2010} derives the stellar baryon fraction within
\rtwo\ rather than \rfive, and derives a relation between stellar mass
and \mtwo.  This choice of aperture clearly provides a more complete
census of the stellar baryon content. It also mitigates the impact of
not including the ICL, leading to a steeper slope that is consistent
with the 
slope we derive when including the ICL. Unfortunately, use of this
radius means that the normalization is not directly comparable with
our studies and others, which select \rfive\ as a radius within which
the gas content can also be accurately measured.  We still include the
\citet{andreon2010} relation in Fig. \ref{fig:comp}, but emphasize
this key difference relative to other studies.

Third, each of the studies listed uses a different \mlrat\ (or in the
case of \citet{leauthaud2012} a distribution of \mlrat\ values) and is
based upon data sets at different redshifts. As discussed above, the
conversion from luminosity to stellar mass is easily the dominant
source of systematic uncertainty.

We make a best attempt in Figure \ref{fig:comp} to compare the various
studies on relatively equal footing. In this Figure we show the
literature results for a \citet{salpeter1955} IMF for those studies
based upon stellar population models. In the case of \citet{lin2012},
which is based upon a \citet{kroupa2001} IMF, we use the prescription
in their paper (multiplying by a factor of 1.34) to convert to a
Salpeter IMF.
Those studies that use dynamical \mlrat\ values are plotted as
published \citep{lin2003,gonzalez2007,andreon2010}. We also plot the
band derived from the extended HOD formalism in Figure 5 of
\citet{leauthaud2012} for a Salpeter IMF, where the width of the band
reflects associated systematic uncertainties. 
We note that the stellar masses used by \citet{giodini2009} have subsequently
been found to be biased high \citep{leauthaud2012,giodini2012},
contributing to the relatively high normalization of that relation.

While there are several relations that are outliers relative to the
ensemble, we recover a normalization and slope that approximate the
median of the literature values if we exclude the fraction of the ICL
that is expected to be missed by these studies. Our total stellar
baryon fraction relation \emph{including} the ICL remains similar in
amplitude to the relation in \ptwo. When we add a deprojection
correction to the stellar mass, which was not done in \ptwo, the
resultant normalization is similar to that of the ``50 kpc'' relation
plotted in Figure \ref{fig:comp}, which excludes the contribution of
the ICL beyond 50 kpc from the BCG --- the ICL contribution and
deprojection correction are roughly equal and opposite in amplitude.

It is clear that the \citet{leauthaud2012} extended HOD results are
significantly lower than direct $f_\star$ determinations, particularly
at low mass where the inclusion of the ICL further exacerbates this
tension (though at even lower mass scales they are consistent with the
direct measurements in \citet{leauthaud2012}).  This fact was
discussed in \citet{leauthaud2012} in the context of a comparison with
\ptwo\ and \citet{giodini2009}.  \citet{leauthaud2012} suggested that
the offset between their work and the \ptwo\ results is due to use of
an excessively large \mlrat\ in \ptwo.  In this paper we have
carefully recalculated the dynamical \mlrat\ values, reconsidering all
assumptions to assess the level of remaining systematic
uncertainties. While the value of \mlrat\ used in this paper is 26\%
lower than in \ptwo\, the offset with \citet{leauthaud2012} remains
despite our efforts to place all studies on equal footing. The reason
for this tension is unclear, but certainly warrants further
investigation.

\section{Conclusions}
\label{sec:summary}

The central goal of this paper is to extend the main
results from \citet{gonzalez2007} using a sample for which we possess
total, gas, and stellar mass measurements for each cluster. The
combination of these measurements for individual systems addresses the
two main weaknesses of our previous analysis: the lack of robust
cluster mass estimates used to derive the stellar mass scaling
relation and the use of disjoint data sets for the gas and stellar
mass measurements. As a result, 
the total baryon fractions in \ptwo\ were based upon scaling relations
rather than derived for individual systems.  Here we present an
analysis of the total baryon fraction within \rfive\ for a sample of
clusters with complete data on the intracluster medium and stellar
component.  These data also enable a direct comparison of the
partitioning of baryons between gas and stars for individual systems
and for stars between galaxies and the brightest cluster galaxy plus intracluster light (BCG+ICL) component.  Our central findings
are:

\begin{enumerate}

\item{We confirm the trend of steeply decreasing stellar mass fraction
    with cluster mass seen in \ptwo, which has now been verified by multiple groups
    for the cluster galaxy population without inclusion of
    intracluster light \citep[$-0.3\ga \alpha \ga-0.55$, Table
    \ref{tab:slopes};][]{giodini2009,andreon2010,zhang2011,lin2012}.
    Our best fit to the current data has a scaling
    $f_{\star,2D}\propto M_{500}^{-0.45\pm0.04}$, including the BCG+ICL;
    if we exclude the fraction of the ICL that we expect is typically
    missed by other studies, then we recover $f_{\star,2D}\propto
    M_{500}^{-0.38\pm0.05}$ with a 26\% lower normalization at
    $10^{14}$\msun\ (Figure \ref{fig:comp}).  The current data are
    less consistent with a simple power-law fit with low scatter than
    our previous data. In particular, the stellar baryon fractions at
    $M<2\times10^{14}$ \msun\ on average are lower than expected from
    a simple extrapolation of the trend observed at higher mass. The
    limited existing data however make any firm conclusions premature. 
    We also recover a scaling relation for the hot gas, $f_{gas}\propto M_{500}^{0.26\pm0.03}$,
    with a slope similar to that of the relation used in our previous work and consistent
    with other determinations in the literature.}

\item{The combination of brightest cluster galaxy plus intracluster
    light is confirmed to contain an decreasing fraction of the total
    luminosity within \rfive\ with increasing cluster mass. For
    systems with \mfive$\approx10^{14}$ \msun, the BCG+ICL together
    typically contribute 40-50\% of the total luminosity within
    \rfive. As discussed in \ptwo, the decrease in the importance of
    the BCG+ICL, and particularly the ICL, with increasing cluster
    mass is consistent with less efficient tidal stripping and
    disruption of galaxies in the more massive systems. We do note
    that one of the \mfive$\approx10^{14}$ \msun\ clusters from
    \citet{sanderson2013} has a lower BCG+ICL luminosity fraction that
    is similar to more massive systems. This system suggests that
    there may be significant scatter in the BCG+ICL luminosity
    fraction at this mass scale, but a larger sample will be required
    to clarify this point.}

\item{The partitioning of baryons between stars and gas within \rfive,
    a measure of the integrated efficiency with which baryons are
    converted to stars,  
    is strongly correlated with \mfive. Our results for the galaxy
    population, excluding ICL beyond 50 kpc from the BCG, yield a
    relation, $M_{\star,2D}/M_{gas}\propto
    M_{500}^{-0.79\pm0.05}$. This power law slope is 2.5$\sigma$ steeper
    than that found by \citet{zhang2011}. While differences in the
    X-ray analyses may contribute, we cannot identify any single
    reason for this $2.5\sigma$ difference. Inclusion of the entire
    contribution from the ICL, which constitutes a greater fraction of
    the stellar mass at low cluster mass, and applying a deprojection
    correction both further steepen this slope, yields
    $M_{\star,3D}/M_{gas}\propto M_{500}^{-0.82\pm 0.05}$. If the
    observed trend continues to lower mass, then
    the stellar contribution may be $>50$\% on mass scales $M_{500}\la
    3\times10^{13}$ \msun.}

\item{The derived relations for $M_{\star,2D}/M_{gas}$ and
    $M_{\star,3D}/M_{gas}$ versus \mfive\ 
    demonstrate a stronger dependence of star formation
    efficiency upon cluster mass than the most recent generation of
    numerical simulations are able to reproduce
    \citep[e.g.,][]{puchwein2010,young2011}. 
    For the $M_{\star,3d}/M_{gas} -$ \mfive\ relation we
    derive an intrinsic scatter of only 12\%.
    This low scatter in the stellar-to-gas mass relation implies
    that the star formation efficiency at a fixed cluster mass scale is quite  
    uniform. It further argues that in
    the $f_*-$\mfive\ relation any departure from a power law at low mass
    must be attributed to either the \mfive\ determinations or changes in
    the total baryon content
    rather than to variations in the star formation efficiency.
    }

\item{ For systems with $M_{500}\ge2\times10^{14}$ \msun, the average
    total baryon fraction is $f_{bary}=0.136\pm0.005$, 18\% below the
    Universal value  for the WMAP7 cosmology. Use of the
    Planck cosmological parameters raises the derived total baryon fraction
    to $f_{bary}=0.144\pm0.005$ for these systems, which at only 7\% below
    the Universal baryon fraction is consistent with this Universal value to within the
    current systematic uncertainties 
    For this Planck cosmology
    there are essentially no missing baryons within \rfive\ for massive galaxy clusters.
  In \ptwo, we concluded that the total baryon fraction was consistent
  with being independent of halo mass on group to cluster scales.
  With the present, improved data, we now see a modest increase with
  mass of the total baryon fraction, $f_{bary}\propto
  M_{500}^{0.16\pm0.04}$.  Below $2\times10^{14}$ \msun, our results
  suggest an increasing physical spread in the total baryon fraction
  among systems \citep[see also][]{sanderson2013}, with fractions
  ranging from 60-90\% (65-100\%) of the WMAP7 (Planck) Universal value.
  Such a variation could arise in a number of ways,
  including variance in the initial conditions or redistribution of
  baryons to beyond \rfive.  However, the relative tightness of the
  $M_{\star}/M_{gas}$ relation, even at these lower halo masses,
  suggests that this scatter must arise in a fashion that affects both
  the stellar and gaseous components equally, arguing against
  late-time hydrodynamic redistribution (e.g., via feedback) of some
  baryons beyond \rfive.}

\item{
Uncertainties associated with the conversion of luminosity
to stellar mass, and with the X-ray derived \mfive, are the most important
sources of systematic uncertainty for the derived stellar baryon fractions.
The assorted systematics associated with the conversion to stellar mass
are present at the level of $\sim15$\%.
     We advocate use of dynamical
    estimates for \mlrat\ due to the minimal required assumptions, and
    advise caution in interpreting results that rely on \mlrat\ values
    derived using stellar population models, for which the choice of
    IMF can change the inferred stellar mass by nearly a factor of two
    \citep[e.g.,][]{leauthaud2012}. Additionally, the recent
    \citet{cappellari2011,cappellari2013} results suggest variation in the shape of
    the initial mass function, and hence \mlrat, with galaxy mass. If
    correct, then current analyses that presume a single IMF at all
    mass scales will yield inherently biased estimates of \mlrat. 

    For \mfive\ both instrumental calibration and departures from
    hydrostatic equilibrium are potential sources of bias. We note that our derived
    \mfive\ are systematically lower than those from \cite{vikhlinin2006} by 8\%, providing
    an approximate estimate of the magnitude of calibration uncertainty. As discussed in
    the text, literature comparisons also indicate that weak lensing \mfive\ determinations may be up to 15\%
    higher than X-ray values. The net impact of a bias of this magnitude would be to decrease $f_{gas}$ and
    $f_{\star}$ by $\sim8$\% and $\sim3$\% respectively, thus decreasing the total baryon fraction by a comparable
    amount. All other quantitative conclusions in this paper, including the slopes in the gas, stellar, and baryon fraction
    relations,  are robust to this level of uncertainty in the total mass.
    }

\end{enumerate}

Given that our new data set provides the most complete census to date
of the hot and cold baryons, the central remaining questions are the
detailed properties of lower mass groups and the redshift evolution of
the baryonic components.  Sanderson et al. (2012) provides a first
step towards addressing the former, while we are using the capability
of the \hst/\wfc\ camera IR channel in conjunction with X-ray data
from \chandra\ and \xmm\ to explore the redshift evolution (programs
12575 and 12634) .

\acknowledgements The authors thank Alastair Sanderson, Trevor Ponman,
Alexie Leauthaud, and Kevin Bundy for extensive discussions pertaining
to this project, and are also grateful to Alexie Leauthaud and Yu-Ying
Zhang for providing data from their papers. We greatly appreciate the
efforts of the anonymous referee, whose suggestions resulted in an
improved paper.  AHG thanks Carnegie Observatories and IPAC for their
hospitality while working on this paper, and acknowledges support from
the National Science Foundation through grant NSF-1108957. SS
acknowledges the Dunlap Institute for Astronomy and Astrophysics for
funding him through the Dunlap Fellowship program.  AIZ and DZ thank
the Max-Planck-Institut f\"ur Astronomie and the Center for Cosmology
and Particle Physics at NYU for their hospitality and support.  AIZ
also acknowledges support through NASA grants NNX08AX81G and
NNX08AC68G associated with XMM observations.

\bibliographystyle{hapj}
\bibliography{ms.bbl}

\appendix

The Planck first year results were released shortly after submission
of this paper \citep{planckcosmology2013}.  In the paper we discuss
the impact on our results of changing from a WMAP7 to Planck
cosmology, but retain WMAP7 as the fiducial cosmological model. In
this Appendix we also provide versions of the main Tables from the
paper for the Planck cosmological parameters.

\newpage
\placetable{tab:planckbaryonfracs}

\begin{deluxetable}{lccc}
\tabletypesize{\scriptsize}
\tablewidth{0pt}
\tablecaption{Derived Mass Fractions for Planck Cosmology ($r<$\rfive)}
\tablehead{
\colhead{Cluster}  & \colhead{$f_{gas}$} & \colhead{$f_{stellar}$} &  \colhead{$f_{baryons}$}\\
\colhead{       }  & \colhead{         } & \colhead{             }  & \colhead{             } 
 }
\startdata
Abell 0122 & $0.094\pm.013$ & $0.026\pm.003$ & $0.120\pm.013$ \\
Abell 1651 & $0.139\pm.013$ & $0.012\pm.001$ & $0.151\pm.013$ \\
Abell 2401 & $0.095\pm.014$ & $0.026\pm.003$ & $0.121\pm.014$ \\
Abell 2721 & $0.134\pm.021$ & $0.016\pm.002$ & $0.150\pm.021$ \\
Abell 2811 & $0.132\pm.011$ & $0.013\pm.002$ & $0.145\pm.012$ \\
Abell 2955 & $0.071\pm.010$ & $0.031\pm.004$ & $0.103\pm.011$ \\
Abell 2984 & $0.117\pm.015$ & $0.041\pm.005$ & $0.159\pm.016$ \\
Abell 3112 & $0.142\pm.010$ & $0.022\pm.002$ & $0.163\pm.010$ \\
Abell 3693 & $0.117\pm.014$ & $0.024\pm.003$ & $0.141\pm.014$ \\
Abell 4010 & $0.127\pm.011$ & $0.023\pm.003$ & $0.150\pm.012$ \\
Abell S0084& $0.094\pm.012$ & $0.024\pm.003$ & $0.118\pm.011$ \\
Abell S0296& $0.081\pm.013$ & $0.020\pm.003$ & $0.101\pm.014$ \\
\hline
Abell 0478 & $0.185\pm.017$ &  ---            & ---             \\
Abell 2029 & $0.139\pm.010$ &  ---            & ---             \\
Abell 2390 & $0.153\pm.025$ &  ---            & ---             \\
\enddata
\tablecomments{The quoted stellar baryon fractions include a deprojection correction, as discussed in the text.}
\label{tab:planckbaryonfracs}
\end{deluxetable}

\begin{deluxetable}{lccccccccc}
\tabletypesize{\scriptsize}
\tablewidth{0pt}
\tablecaption{Observed Cluster Properties for Planck Cosmology}
\tablehead{
\colhead{Cluster} &\colhead{$z$} &  $T_{X,2}$ & \colhead{$L_{BCG+ICL}$} & \colhead{$L_{Total}$}  &\colhead {\rfive}           & \colhead {\mfive} & \colhead {$M_{gas,500}$}       & \colhead {$M_{\star,2D,500}$} & \colhead{$M_{\star,3D,500}$} \\
\colhead{       } &\colhead{  }  &   \colhead{(keV)} & \colhead{($10^{12}$\lsun)}         & \colhead{($10^{12}$\lsun)}        &\colhead {(Mpc)} & \colhead {($10^{14}$ \msun)}   & \colhead {($10^{13}$ \msun)} & \colhead {($10^{13}$ \msun)} & \colhead{($10^{13}$ \msun)}
 }
\startdata
Abell 0122 & 0.1134 &  $3.65\pm0.15$     & $0.91\pm0.04$ & $ 2.87\pm0.18 $ & $0.93\pm .03$ & $2.35\pm.20$ &  $2.20\pm .23$ & $0.76\pm .05$ & $0.61\pm .04$ \\
Abell 1651 & 0.0845 &  $6.10\pm0.25$     & $0.93\pm0.09$ & $ 3.17\pm0.23 $ & $1.23\pm .03$ & $5.37\pm.43$ &  $7.45\pm .35$ & $0.84\pm .06$ & $0.67\pm .05$ \\
Abell 2401 & 0.0571 &  $2.06\pm0.07$     & $0.35\pm0.01$ & $ 1.23\pm0.08 $ & $0.71\pm .03$ & $0.99\pm.11$ &  $0.95\pm .10$ & $0.33\pm .02$ & $0.26\pm .02$ \\
Abell 2721 & 0.1144 &  $4.78\pm0.23$     & $0.61\pm0.01$ & $ 2.82\pm0.20 $ & $1.07\pm .03$ & $3.60\pm.33$ &  $4.82\pm .63$ & $0.75\pm .05$ & $0.58\pm .04$ \\
Abell 2811 & 0.1079 &  $4.89\pm0.20$     & $0.91\pm0.15$ & $ 2.12\pm0.18 $ & $1.08\pm .03$ & $3.73\pm.29$ &  $4.94\pm .19$ & $0.56\pm .05$ & $0.47\pm .04$ \\
Abell 2955 & 0.0943 &  $2.13\pm0.10$     & $0.64\pm0.04$ & $ 1.46\pm0.08 $ & $0.71\pm .03$ & $1.03\pm.12$ &  $0.74\pm .05$ & $0.39\pm .02$ & $0.32\pm .02$ \\
Abell 2984 & 0.1042 &  $2.08\pm0.07$     & $0.92\pm0.03$ & $ 1.79\pm0.09 $ & $0.70\pm .02$ & $0.99\pm.10$ &  $1.16\pm .09$ & $0.48\pm .02$ & $0.41\pm .02$ \\
Abell 3112 & 0.0750 &  $4.54\pm0.11$     & $0.98\pm0.05$ & $ 3.45\pm0.23 $ & $1.06\pm .02$ & $3.37\pm.20$ &  $4.77\pm .18$ & $0.92\pm .06$ & $0.73\pm .04$ \\
Abell 3693 & 0.1237 &  $3.63\pm0.20$     & $0.76\pm0.08$ & $ 2.74\pm0.20 $ & $0.93\pm .03$ & $2.36\pm.24$ &  $2.76\pm .16$ & $0.73\pm .05$ & $0.57\pm .04$ \\
Abell 4010 & 0.0963 &  $3.78\pm0.13$     & $0.87\pm0.13$ & $ 2.68\pm0.21 $ & $0.95\pm .02$ & $2.51\pm.19$ &  $3.18\pm .13$ & $0.71\pm .06$ & $0.57\pm .05$ \\
Abell S0084& 0.1100 &  $3.75\pm0.20$     & $0.77\pm0.03$ & $ 2.78\pm0.19 $ & $0.94\pm .03$ & $2.47\pm.25$ &  $2.32\pm .17$ & $0.74\pm .05$ & $0.58\pm .04$ \\
Abell S0296& 0.0696 &  $2.70\pm0.21$     & $0.61\pm0.02$ & $ 1.38\pm0.08 $ & $0.81\pm .04$ & $1.51\pm.21$ &  $1.22\pm .11$ & $0.37\pm .02$ & $0.31\pm .01$ \\
\hline
Abell 0478 & 0.0881 &  $7.09\pm0.12$     &  ---            &  ---             & $1.33\pm 0.03$ & $6.80\pm0.39$ &  $12.6\pm 0.9$ & ---  &  --- \\
Abell 2029 & 0.0773 &  $8.41\pm0.12$	    &  ---            &  ---             & $1.47\pm 0.03$ & $9.01\pm0.57$ &  $12.5\pm 0.4$ & --- &   --- \\
Abell 2390 & 0.2329 &  $10.6\pm0.8$	    &  ---            &  ---             & $1.53\pm 0.08$ & $12.1\pm1.8$ &  $18.5\pm 1.2$ & --- &   --- \\
\enddata
\enddata
\tablecomments{Abell 0478, Abell 2029, and Abell 2390 are not part of the main sample. These clusters were included only in the X-ray analysis to extend the baseline to higher mass, but have no photometry equivalent to the other systems with which to measure the stellar mass. At the high mass end; however, the stellar component contributes a relatively small fraction of the total baryons. The luminosities include appropriate e+k corrections for each galaxy from \ptwo. The stellar masses are quoted
as observed, with no deprojection correction applied. }
\label{tab:planckdata}
\end{deluxetable}

\begin{deluxetable}{lcr}
\tabletypesize{\scriptsize}
\tablewidth{0pt}
\tablecaption{Derived $M_j-$\mfive\ Relations for Planck Cosmology}
\tablehead{
\colhead{Component}  & \colhead{$a$} & \colhead{$b$}
 }
\startdata
$M_{\star,2D}$& $3.9\pm0.2 \times10^{-2}$ &  $0.56\pm 0.05$\\	
$M_{\star,3D}$ & $3.3\pm0.2 \times10^{-2}$ &  $0.53\pm 0.05$\\	
$M_{gas}$     & $ 9.2\pm0.4\times10^{-2}$ &  $1.26\pm0.03$\\	
$M_{bary}$   & $1.21\pm0.05\times10^{-1}$ &  $1.17\pm0.04$\\
\hline
$M_{\star,2D}/M_{gas}$  & $4.70\pm0.03\times10^{-1}$ &  $-0.85\pm0.05$\\
$M_{\star,3D}/M_{gas}$  & $3.87\pm0.02\times10^{-1}$ &  $-0.81\pm0.05$\\
\enddata
\tablecomments{ Best fit parameters for the relation $M_j=a (M_{500}/10^{14}M_\odot)^b$, where $M_j$ is the mass contained in each baryonic component. $M_{bary}$ is derived using the deprojected stellar mass. The slope for the baryon fraction relations is equivalent to $1-b$. We also include the best fit parameters for the stellar-to-gas mass ratios as a function of \mfive. The relation for the gas mass is derived including the clusters from \citet{vikhlinin2006}, while the other relations are derived using only clusters with both gas and stellar mass data.}
\label{tab:planckbestfit}
\end{deluxetable}

\end{document}